\documentclass[eqsecnum,aps,showpacs,amsmath,twocolumn,superscriptaddress]{revtex4}
\usepackage{epsfig}
\usepackage{color}
\usepackage{graphicx}
\usepackage{amssymb}

\begin{document}
%\preprint{20170524 Shinkai}
\title{Nonlinear dynamics in the Einstein-Gauss-Bonnet gravity\footnote{Accepted for publication in Physical Review D. (2017)}}
%\date{June 7, 2017}
\date{\today}

\author{Hisa-aki Shinkai}\email{hisaaki.shinkai@oit.ac.jp}
\affiliation{Department of Information Systems, Faculty of Information Science \& Technology, 
Osaka Institute of Technology,
Kitayama, Hirakata City, Osaka 573-0196, Japan}
\author{Takashi Torii}\email{takashi.torii@oit.ac.jp}
\affiliation{Department of System Design, Faculty of Robotics \& Design, 
Osaka Institute of Technology, 
Kita-ku, Osaka City,  Osaka 530-8568, Japan}

\begin{abstract}
We numerically investigated how the nonlinear dynamics depends on the dimensionality and on the higher-order curvature corrections in the form of Gauss-Bonnet (GB) terms.  We especially monitored the processes of appearances of a singularity (or black hole) in two models: (i) a perturbed wormhole throat in spherically symmetric space-time, and (ii) colliding scalar pulses in plane-symmetric space-time. 
We used a dual-null formulation for evolving the field equations, which enables us to locate the trapping horizons directly, and also enables us to follow close to the large-curvature region due to its causal integrating scheme.
We observed that the fate of a perturbed wormhole is either a black hole or an expanding throat depending on the total energy of the structure, and its threshold depends on the coupling constant of the GB terms ($\alpha_{\rm GB}$).  
We also observed that a collision of large scalar pulses will produce a large-curvature region, of which the magnitude also depends on $\alpha_{\rm GB}$. For both models, the normal corrections ($\alpha_{\rm GB}>0$) work for avoiding the appearance of singularity, although it is inevitable. We also found that in the critical situation for forming a black hole, the existence of the trapped region in the Einstein-GB gravity does not directly indicate the formation of a black hole.
\end{abstract}
%<<<<<<<<<<<<< PACS NUMBER >>>>>>>>>>>>>>>%
\pacs{04.20.-q, 04.40.-b, 04.50.-h}
% 04.20.-q : Classical general relativity
% 04.40.-b : Self-gravitating systems; continuous media and classical fields in curved spacetime
% 04.50.-h : Higher-dimensional gravity and other theories of gravity

%######################################################################%
\maketitle
%######################################################################%

%######################################################################%
%######################################################################%
%    SECTION  1 
%######################################################################%
%######################################################################%
\section{Introduction}
%======================================%
%\subsection{Einstein-Gauss-Bonnet gravity}
%======================================%
Nobody raises an objection to the fact that general relativity (GR) describes the nature of strong gravity quite well.  The success of the standard big-bang theory is recognized as the most successful physical result in the 20th century, and the black-hole physics is now applied to understand several field theories and/or material physics. We have also seen the first direct detection of gravitational wave, achieved a century after Einstein's theoretical discovery. 

One of our most exciting topics now is what the physics laws beyond GR are. We know that GR cannot merge with quantum theory in its current form.  We also know that standard cosmology still requires new ideas to explain the matter contents and the rate of expansion of space-time.  There are several approaches to these problems.  
Among them, we think that gravity theories in higher-dimensional space-time and/or in the theories with higher-order curvature terms are the natural extensions to be considered. 

We present in this article  several nonlinear behaviors in gravity theory with the Gauss-Bonnet (GB) terms \cite{Lovelock1971,Gross,Tseytlin}. 
The Einstein-GB gravity is derived from string theory, with additional  
higher-order-curvature correction terms to GR in the form of the Lagrangian, 
%Equation--------------------------%
\begin{eqnarray}
{\cal L}_{\rm GB}={\cal R}^2-4{\cal R}_{\mu\nu}{\cal R}^{\mu\nu}
+{\cal R}_{\mu\nu\rho\sigma}{\cal R}^{\mu\nu\rho\sigma},   \label{GBaction}
\end{eqnarray}
where ${\cal R}$,
${\cal R}_{\mu\nu}$, and ${\cal R}_{\mu\nu\rho\sigma}$ are 
 the $n$-dimensional scalar curvature, the Ricci tensor, and the Riemann curvature, respectively. 
%--------------------------------------%
This particular combination gives us 
several reasonable properties,  as such ghost-free combinations \cite{Zwiebach}, 
and a set of equations up to the second derivative in spite of the higher-curvature combinations. 
The theory is expected to have singularity-avoidance features in the context of
gravitational collapses and/or cosmology.  
However, only a few studies so far have reported on the investigation of nonlinear dynamical features in 
Einstein-GB gravity  
(e.g., numerical studies on critical phenomena \cite{GolodPiran,Deppe2012}, 
black hole formation in AdS \cite{Deppe2014PRL,Deppe2016JHEP}).
% are recently reported for small coupling constant, $\alpha_{\rm GB}$ ). %Izaurieta,

%======================================%
%\subsection{Wormhole dynamics}
%======================================%
Our first investigative model concerns wormhole dynamics. 
A wormhole is a hypothetical object such as 
a short-cut tunnel connecting two points in space-time. 
The idea is frequently used in science fiction to allow for rapid interstellar travel, warp drives, and time machines.  
However, wormholes are also a theoretical research topic 
with a long history (See a review, e.g. Visser \cite{MVbook} for earlier works; see also e.g., Lobo
\cite{Lobobook} \& \cite{WHbook} for recent works.)

We are especially interested in the fate of a perturbed Ellis wormhole \cite{Ellis}, 
whose behavior is well known in four-dimensional GR. 
The Ellis wormhole is constructed with a massless Klein-Gordon field whose kinetic term
takes the sign opposite to normal, which was rediscovered by Morris and  Thorne \cite{MT},  who 
considered ``traversable conditions" for human travel 
through wormholes in a response to Carl Sagan's idea for his novel {\it Contact}. 

The first numerical simulation on its 
stability behavior was reported by one of the authors \cite{ShinkaiHayward}.
It shows that the Ellis wormhole is unstable against the injection of perturbed field to the throat, and 
the wormhole will be changed either to a black hole
 or to an expanding throat depending on the energy balance. 
These basic behaviors were repeatedly confirmed by other groups
\cite{Doroshkevich,GonzalezGuzmanSarbach2009cqg1,GonzalezGuzmanSarbach2009cqg2,GonzalezGuzmanSarbach2009prd}. 
We will explain in more detail in Sec. \ref{section3}. 

In this article, we present numerical evolutions of higher-dimensional wormholes with the GB terms. 
Wormhole studies in higher-dimensional space-time is not a new topic. 
We can find articles on the subject from the 1980s \cite{ChadosDetweiler,Clement84}, and 
recent studies are including higher-curvature terms 
(see,  e.g.,  \cite{MaedaNozawa}, Refs. \cite{KantiKleihausKunz2011,KantiKleihausKunz2012},  and references therein).
Most of the research mainly concerns the solutions and their energy conditions, 
but to our knowledge there is no general discussion on the nonlinear stability issues of the
solutions (linear stability analysis can be found in Refs. \cite{KantiKleihausKunz2011,KantiKleihausKunz2012}). 
Studies on wormholes in Einstein-GB gravity have long histories. 
Several solutions and their classifications are reported in Refs. \cite{BhawalKar,Dotti}, 
while their energy conditions are considered in Ref. \cite{MaedaNozawa}.
Similar research is extended to Lovelock gravity \cite{DehghaniDayyani}, 
%to the Yang-Mills-Gauss-Bonnet system \cite{Mazharimousavi}, 
and also to the dilatonic GB system \cite{KantiKleihausKunz2011,KantiKleihausKunz2012}.
%Our aim is to investigate their dynamical features.  

A couple of years ago,  we constructed Ellis-type solutions in higher-dimensional GR 
and reported stability analysis using a linear perturbation method \cite{ToriiShinkaiWH}.  
The solutions have at least one negative mode, which leads to the conclusion that 
all Ellis-type   (static and spherically symmetric)   wormholes in GR are linearly unstable
\footnote{
  Recent studies 
\cite{MatosNunes2006,DzhunushalievFolomeevKleihausKunz2013,KleihausKunz2014} of rotating 
Ellis-type wormholes show that this unstable mode is no longer present.   
}.
The time scale of instability becomes shorter as $n$ becomes large. 
Therefore, the confirmation of these predictions and the dynamical behavior 
with the GB terms are two main objectives in Sec. \ref{section3}.

%======================================%
%\subsection{Singularity formation by colliding waves}
%======================================%
Our second investigative model deals with colliding wave packets.
Due to the nonlinear features of the theory, in GR, 
gravitational waves interact with themselves when they pass through
each other.
Considering a collision of plane gravitational waves is the
 simplest scenario of this nonlinear interaction problem
 (see Ref.~\cite{GriffithsBook} and references therein).

In fact, Penrose \cite{Penrose1965} pointed out that the future light cone of 
 a plane wave is distorted as it passes through another plane wave. 
As one aspect of this global property, 
 Szekeres \cite{Szekeres19701972} and Khan and Penrose \cite{KhanPenrose1971}
found exact solutions of colliding plane waves 
in flat space-time, which form a curvature singularity
in their interacting region.
Stewart {\it et al.} \cite{StewartFriedrich1982,CorkillStewart1983} performed numerical
simulations in the framework of a 2+2 decomposition of space-time, and
found that the expansion of the null geodesic will be negative
after a collision of waves. 
Since these solutions assume a plane-symmetric space-time, this
singularity does not have a horizon; it is a ``naked" one. 

Our attention to this problem focuses on the differences in the growth of curvature, 
especially the dependences on the dimension and the GB terms.  
We have found that we can compare the behaviors more easily when we place colliding 
matter rather than colliding gravitational waves. Therefore, we prepare the model of colliding normal scalar packets in plane-symmetric space-time, and we show comparisons in Sec. \ref{section4}.

~

The construction of this article is as follows. 
In Sec. \ref{section2}, we 
show the set of field equations in the form of a dual-null coordinate system and will explain our numerical schemes.  
We then show the results of the evolutions of a perturbed wormhole in Sec. \ref{section3}, 
and the results of the collision of scalar plane pulses in Sec. \ref{section4}. 
Section \ref{section5}  provides a  summary. 

%######################################################################%
%######################################################################%
%    SECTION  2 
%######################################################################%
%######################################################################%
\section{Field Equations and numerical technique}\label{section2}
%%%%%%%%%%%%%%%%%%%%%%%%%%%%%%%%%%%%
\subsection{Action}\label{sec_GaussBonnet}
%%%%%%%%%%%%%%%%%%%%%%%%%%%%%%%%%%%%

The Einstein-GB action in 
$n$-dimensional space-time $({\cal M}, g_{\mu\nu})$ 
is described as
%Equation--------------------------%
\begin{eqnarray}
\label{bulk_action}
S &=&
\int_{\cal M} d^{n}x \sqrt{- g} \biggl[ 
\frac{1}{2 \kappa^2} \bigl(\alpha_{\rm GR} {\cal R} 
-  2\Lambda 
+\alpha_{\rm GB} {\cal L}_{\rm GB}\bigr) \nonumber \\&& \qquad \qquad\qquad
+{\cal L}_{\rm matter} \biggr],
\end{eqnarray}
%--------------------------------------%
%\end{widetext}
where ${\cal L}_{\rm GB}$ is the GB term [Eq. (\ref{GBaction})], 
$\kappa^2$ is the $n$-dimensional gravitational constant, and ${\cal L}_{\rm matter}$ is the matter Lagrangian. This action reproduces the standard $n$-dimensional Einstein gravity, if we set the coupling constant $\alpha_{\rm GB}$ equal to zero. 
On the other hand, by setting $\alpha_{\rm GR}=0$, the system becomes pure GB gravity.
In the actual simulations, we set $\alpha_{\rm GR}=1$, $\Lambda=0$, $\kappa^2=1$ and change $\alpha_{\rm GB}$ as a parameter while we write the set of equations with $\alpha_{\rm GR}$ and $\Lambda$ in this section in order 
to compare the terms with those from ${\cal L}_{\rm GB}$. 

The action (\ref{bulk_action}) gives the gravitational equation as
%Equation--------------------------%
\begin{eqnarray}
\alpha_{\rm GR} {G}_{\mu\nu}+g_{\mu\nu} \Lambda
+ \alpha_{\rm GB} {H}_{\mu\nu}  = \kappa^2 \,{T}_{\mu\nu} \, ,
\label{EinsteinGB}
\end{eqnarray}
%--------------------------------------%
where 
%Equation--------------------------%
\begin{eqnarray}
{G}_{\mu\nu}=&& \!\!\!\!\!{\cal R}_{\mu\nu}-\frac{1}{2}g_{\mu\nu}{\cal R},
\\
%--------------------------------------%
{H}_{\mu\nu}=&& \!\!\!\!\!
2\Bigl({\cal R}{\cal R}_{\mu\nu}
-2{\cal R}_{\mu\alpha}{\cal R}^{\alpha}_{~\nu}
-2{\cal R}^{\alpha\beta}{\cal R}_{\mu\alpha\nu\beta}
\nonumber \\ && \;\;\;\;\;\;\;
+{\cal R}_{\mu}^{~\alpha\beta\gamma}{\cal R}_{\nu\alpha\beta\gamma}\Bigr)
-\frac{1}{2}g_{\mu\nu}{\cal L}_{\rm GB},
\label{EGB:eq}
\\ 
%--------------------------------------%
{T}_{\mu\nu} =&& \!\!\!\!\!
%-\frac{\Lambda}{\kappa^2}g_{\mu\nu}
-2 {\delta {\cal L}_{\rm matter} \over \delta
g^{\mu\nu}}  +g_{\mu\nu}{\cal L}_{\rm matter}.
\label{em_tensor_of_bulk}
\end{eqnarray}
%--------------------------------------%

%%%%%%%%%%%%%%%%%%%%%%%%%%%%%%%%%%%%
\subsection{Dual-null formulation}
%%%%%%%%%%%%%%%%%%%%%%%%%%%%%%%%%%%%
We use  dual-null formulation for expressing space-time which has spherical symmetry 
(Sec. \ref{section3})
or  planar symmetry (Sec. \ref{section4}) 
\footnote{We formulated $n+1$ decomposition of the Einstein-GB field equations in Ref. \cite{ToriiShinkai2008}, 
but we decided to apply the dual-null formulation in this study since there are many advantages as we wrote in the text.}. 
The use of dual-null coordinates simplifies the treatment of horizon dynamics, 
enables us to approach close to large-curvature regions, 
and also clarifies radiation propagation in far regions. 
We implement our dual-null evolution code which was used for four-dimensional GR \cite{ShinkaiHayward}
so as to follow higher-dimensional space-time with the GB terms. 

\begin{widetext}
We adopt the line element
\begin{equation}%========<Equation>========%
\label{metric}
ds^2=-2e^{f(x^+\!, \:x^-)}dx^+\, dx^- + r^2(x^+\!, \:x^-)\gamma_{ij}dz^i dz^j,
\end{equation}  %--------------------------%
where the coordinates $(x^+,x^-)$
are along the null propagation directions, 
and $\gamma_{ij}dx^i dx^j$ 
is the metric of the $(n-2)$-dimensional unit constant-curvature space with $k=\pm1,\:0$.
%\footnote{For convenience, we write the metric components, 
%$g_{+-}=g_{-+}=-e^{-f}, g^{+-}=g^{-+}=-e^{f}, g_{ij}=r^2\gamma_{ij}, g^{ij}=r^{-2}\gamma^{ij}$. }

For writing down the field equations, we introduce the variables
\begin{eqnarray}
\Omega&\equiv&\frac{1}{r}, \\
\vartheta_\pm &\equiv& (n-2)\partial_\pm r, \\
\nu_\pm &\equiv& \partial_\pm f, 
\end{eqnarray}
where $\partial_\pm \equiv \partial/\partial x^\pm$,  
and these are the conformal factor, expansions, and inaffinities, respectively. 

%%%%%%%%%%%%%%%%%%%%%%%%%%%%%%%%%%%%
%\subsection{Field equations}
%%%%%%%%%%%%%%%%%%%%%%%%%%%%%%%%%%%%
The nonzero Einstein tensor components, then, are 
\begin{eqnarray}  
\alpha_{\rm GR} G _{ ++}&=& 
-\Omega(\partial_+\vartheta_+ + \vartheta_+ \nu_+)\left( 1+2\tilde{\alpha}\Omega^2 Z\right),
 \\ 
 \alpha_{\rm GR} G _{ --}&=& 
-\Omega(\partial_-\vartheta_- + \vartheta_- \nu_-)\left( 1+2\tilde{\alpha}\Omega^2 Z\right),
%\end{eqnarray}  
%\begin{widetext}
%\begin{eqnarray}  
 \\ 
 \alpha_{\rm GR} G _{+-}&=& 
\Omega \partial_-\vartheta_+ + \frac{(n-2)(n-3)\Omega^2}{2}  \left[ k e^{-f}+\frac{2}{(n-2)^2}\vartheta_+\vartheta_-\right]
\nonumber \\&&
+\tilde{\alpha}\left[ \frac{(n-2)(n-5)}{2}k^2\Omega^4e^{-f}
+2\Omega^3 Z \partial_-\vartheta_+
+\frac{2(n-5)}{n-2}\Omega^4 Z \vartheta_+\vartheta_-\right]- {\Lambda}e^{-f},
\label{eqGuv}
 \\ 
 \alpha_{\rm GR} G _{ ij}&=& 
\gamma_{ij}\left\{
e^f \left[\frac{\partial_{(+} \nu_{-)}}{\Omega^2} 
- \frac{2(n-3)}{(n-2)\Omega}\partial_{(-}\vartheta_{+)}
-\frac{(n-3)(n-4)}{(n-2)^2} \, \vartheta_+\vartheta_-\right]-\frac{(n-3)(n-4)k}{2}
\right\}
\nonumber \\&&
+{\tilde{\alpha}}\gamma_{ij}\left\{
2 e^fZ \left[ \partial_{(+} \nu_{-)} -\frac{2(n-5)\Omega}{n-2} {\partial_{(-}\vartheta_{+)}} -\frac{(n-5)\Omega^2}{n-2} {\vartheta_+} {\vartheta_-}\right] \right.
\nonumber \\&&
~~~~+\frac{4e^{2f}}{(n-2)^2}\bigl[({\partial_+\vartheta_+}+\nu_+ {\vartheta_+}) ({\partial_-\vartheta_-}+\nu_- {\vartheta_-})
-({\partial_{(-}\vartheta_{+)}})^2
\bigr]
\nonumber \\&&
 \left. 
~~~~+2(n-5)Z^2\Omega^2
-\frac{(n-2)(n-5)}{2}k^2\Omega^2 \right\}
+{\Lambda}r^2 \gamma_{ij}, 
\end{eqnarray} 
%\end{widetext}
%
where $\tilde{\alpha}=(n-3)(n-4) \alpha_{\rm GB}$, %$\tilde{\Lambda}=2\Lambda/(n-1)/(n-2)$
$\displaystyle Z=k+W$, 
$\displaystyle W=\frac{2e^f}{(n-2)^2}\vartheta_+\vartheta_-$, and we use the expression,  
$\displaystyle a_{(+}b_{-)}=\frac{1}{2}(a_+b_-+a_-b_+)$. 

The set of dual-null field equations, then, becomes 
%\begin{widetext}
\begin{eqnarray}
\partial_+\vartheta_+&=&-\vartheta_+ \nu_+
-\frac{1}{\Omega A}\kappa^2 T_{++},
\label{eqEinsteinpp}
\\
\partial_-\vartheta_-&=&-\vartheta_- \nu_-
-\frac{1}{\Omega A}\kappa^2 T_{--},
\label{eqEinsteinmm}
\\
\partial_-\vartheta_+&=&
\frac{1}{\Omega A}
\left[ - \frac{\alpha_{\rm GR}(n-2)(n-3)}{2} \Omega^2e^{-f}Z+ e^{-f}\Lambda + \kappa^2 T_{+-} \right]
%\nonumber \\ &&
-\frac{\tilde{\alpha}(n-2)(n-5)}{2}\frac{\Omega^3e^{-f}}{A}\left( k^2 +2W Z\right),
\label{eqEinsteinpm}
\end{eqnarray}
and
\begin{eqnarray}
\partial_+ \nu_-&=&
%%%%%%  new equation line  %%%%%%%%
% line 1
\alpha_{\rm GR}(n-3) \frac{Ze^{-f}\Omega^2}{A}    \left[  
-\frac{\alpha_{\rm GR}(n-3)}{A}
+\frac{(n-4)}{2}
\right]
\nonumber \\ &&
+\frac{e^{-f} \Lambda}{A} \left[
\frac{2\alpha_{\rm GR}(n-3)}{(n-2)A}-1 \right]
+
\frac{2(n-3)}{(n-2)A^2}
\kappa^2 T_{+-} 
+\frac{\Omega^2e^{-f}}{A}\kappa^2 T_{zz}
\nonumber \\&&
% line 2
+\tilde{\alpha}(n-5)\frac{\Omega^2e^{-f}}{ A^2}
\biggl\{ - {\alpha_{\rm GR}}
{(n-3)}\Omega^2 ( k^2 +2W Z +2Z^2) 
-2  \tilde{\alpha} (n-5)\Omega^4 (k^2 +2W Z)Z
\nonumber \\&&
% line 5
%
\qquad \qquad \qquad \qquad 
\left.
+\frac{\Omega^2 A}{2}
\left[{(n-2)}k^2 +2WZ -4Z^2\right]
% line 3
%
 +  \frac{4Z}{n-2} 
(\Lambda + e^f\kappa^2 T_{+-}) \right\}
\nonumber \\&&
% line 6
%
-\frac{4\tilde{\alpha}}{(n-2)^2}\frac{\Omega^2 e^{f}}{ A}
\left[
({\partial_+\vartheta_+}+\nu_+ {\vartheta_+}) ({\partial_-\vartheta_-}+ \nu_- {\vartheta_-})
-({\partial_{(-}\vartheta_{+)}})^2\right] ,
\label{eqEinsteinzz}
\end{eqnarray}
where $A=\alpha_{\rm GR}+2\tilde{\alpha}\Omega^2 Z$.
Note that $\partial_+ \vartheta_-=\partial_- \vartheta_+$ and $\partial_+ \nu_-=\partial_- \nu_+$.  
%and there is no equations for $\partial_+ \nu_+$ and $\partial_- \nu_-$. 

\end{widetext}

~

~

%%%%%%%%%%%%%%%%%%%%%%%%%%%%%%%%%%%%
\subsection{Matter terms}
%%%%%%%%%%%%%%%%%%%%%%%%%%%%%%%%%%%%
We assume two scalar fields, the normal field $\psi(x^+,x^-)$ and the ghost field $\phi(x^+,x^-)$, 
\begin{equation}
T_{\mu\nu}
=T_{\mu\nu}^\psi+T_{\mu\nu}^\phi, 
\end{equation}
{\rm where}
\begin{eqnarray}
T_{\mu\nu}^\psi&=& \partial_\mu \psi \partial_\nu\psi-g_{\mu\nu}\Bigl[\frac{1}{2}(\nabla\psi)^2+V_{\psi}(\psi)\Bigr], \\
T_{\mu\nu}^\psi&=&
-\partial_\mu \phi \partial_\nu\phi-g_{\mu\nu}\biggl[-\frac{1}{2}(\nabla\phi)^2+V_{\phi}(\phi)\biggr], 
\end{eqnarray}
both obey the Klein-Gordon equations, 
\begin{equation}
\Box \psi =\frac{dV_{\psi}}{d\psi}, \qquad \Box\phi = \frac{dV_{\phi}}{d\phi}, 
\label{KGKG}
\end{equation}
respectively. 
If we define the scalar momenta as 
\begin{eqnarray}
\pi_\pm &\equiv& r\partial_\pm\psi 
= \frac{1}{\Omega}\partial_\pm\psi, \\
p_\pm &\equiv& r \partial_\pm\phi 
= \frac{1}{\Omega}\partial_\pm\phi, 
\end{eqnarray}
then the nonzero $T_{\mu\nu}$ components are 
%$(\nabla\phi)^2=-2e^f\phi_u\phi_v$ 
\begin{eqnarray}
T_{++}&=&%\psi_u\psi_u-\phi_u\phi_u =\frac{1}{r^2}(\pi_+^2-p_+^2) = 
{\Omega^2}(\pi_+^2-p_+^2), \label{eqT++}\\
T_{--}&=&%\psi_v\psi_v-\phi_v\phi_v =\frac{1}{r^2}(\pi_-^2-p_-^2)= 
{\Omega^2}(\pi_-^2-p_-^2), \\
T_{+-}&=&T_{-+}=
%\{\psi_u\psi_v+(-\psi_u\psi_v+e^{-f}V_{\psi})\} +\{-\phi_u\phi_v+(\phi_u\phi_v+e^{-f}V_{\phi})\} 
e^{-f}(V_{\psi} + V_{\phi}),
\\
T_{z_iz_j}&=&
\Bigl[e^f(\pi_+\pi_--p_+p_-)-\frac{1}{\Omega^2}(V_{\psi}+V_{\phi})\Bigr] \nonumber \\ && ~~~\times \gamma_{ij}. 
\label{eqTzz}
\end{eqnarray}

Equation ~(\ref{KGKG}) becomes
\begin{eqnarray}
2\partial_+ \pi_- &=& 
\frac{4-n}{n-2}  \Omega \vartheta_+ \pi_- 
-\Omega \vartheta_- \pi_+ 
-\frac{1}{e^f \Omega} \frac{dV_{\psi}}{d\psi},~~~
\label{eq_pip5d}
\\
2\partial_- \pi_+ &=& 
\frac{4-n}{n-2} \Omega \vartheta_- \pi_+ 
-\Omega \vartheta_+ \pi_- 
-\frac{1}{e^f \Omega}  \frac{dV_{\psi}}{d\psi},
\label{eq_pim5d}
\\
2\partial_+ p_- &=& 
\frac{4-n}{n-2} \Omega \vartheta_+ p_- 
-\Omega \vartheta_- p_+ 
-\frac{1}{e^f \Omega}  \frac{dV_{\phi}}{d\phi},
\label{eq_pp5d}
\\
2\partial_- p_+ &=& 
\frac{4-n}{n-2} \Omega \vartheta_- p_+ 
-\Omega \vartheta_+ p_- 
-\frac{1}{e^f \Omega}  \frac{dV_{\phi}}{d\phi}.
\label{eq_pm5d}
\end{eqnarray}
These equations complete the system. 

%%%%%%%%%%%%%%%%%%%%%%%%%%%%%%%%%%%%
\subsection{Numerical integration scheme}
%%%%%%%%%%%%%%%%%%%%%%%%%%%%%%%%%%%%

\begin{figure}[thb]
\begin{center}
\includegraphics[width=14pc]{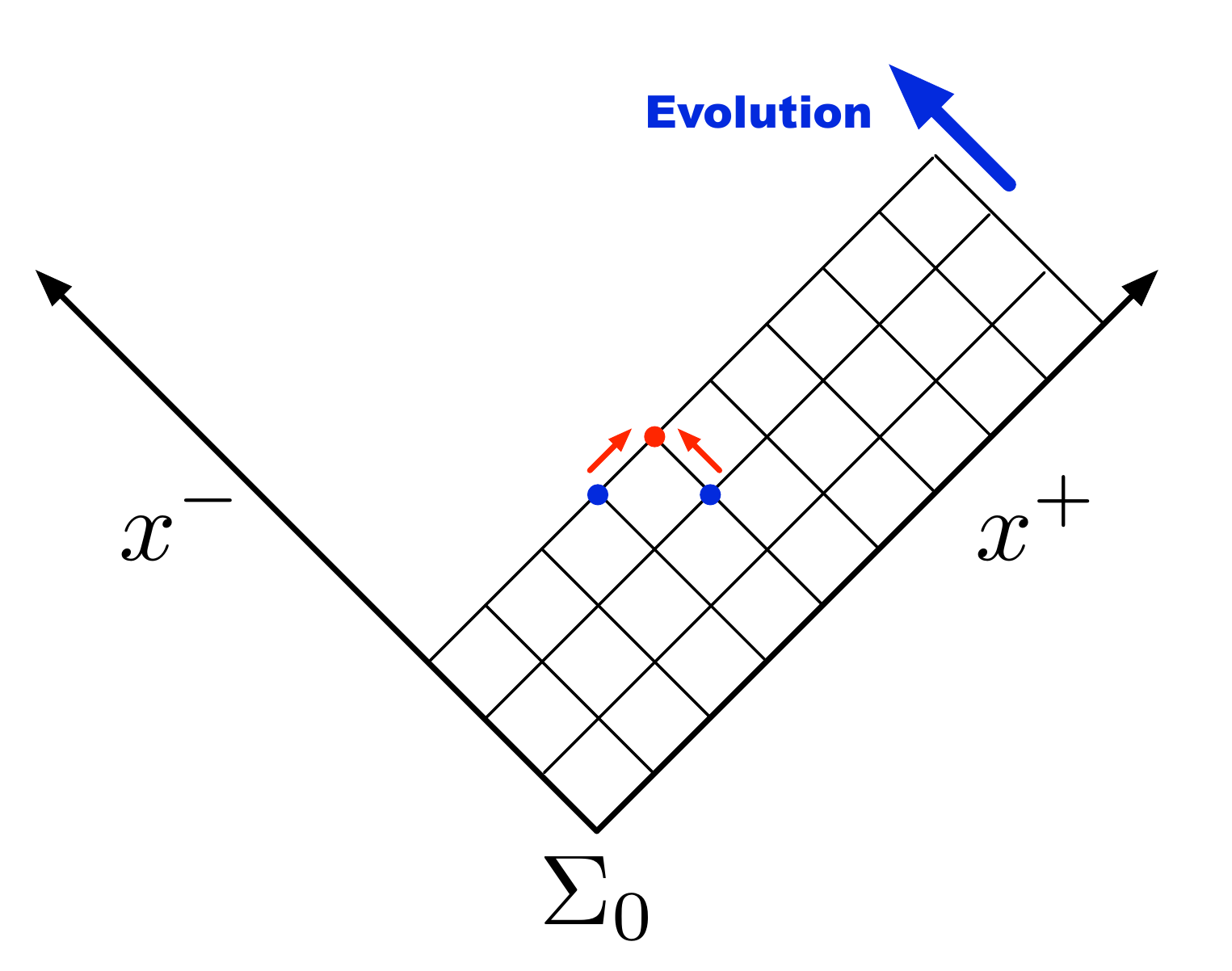}\hspace{2pc}%
\begin{minipage}[b]{20pc}
\caption{\label{fig:2002fig1} Numerical grid structure.  Initial data are given on null
hypersurfaces $\Sigma_\pm$ ($x^\mp=0$, $x^\pm>0$) and their intersection $\Sigma_0$.
}
\end{minipage}
\end{center}
\end{figure}

The basic idea of numerical integration is as follows. 
We prepare our numerical integration range as drawn in Fig.~\ref{fig:2002fig1}. 
We give initial data on a surface $\Sigma_0$, where $x^+=x^-=0$, and the two null hypersurfaces $\Sigma_\pm$, generated from it, where $x^\mp=0$ and $x^\pm>0$.
Generally, the initial data have to be given as
\begin{eqnarray}
&&(\Omega,f,\vartheta_\pm,\phi,\psi)\quad\hbox{on $\Sigma_0$}\\
&&(\nu_\pm,p_\pm,\pi_\pm)\quad\hbox{on $\Sigma_\pm$.}
\end{eqnarray}
We then evolve the data $u=(\Omega, \vartheta_\pm, f, \nu_\pm, \phi,
\psi, p_\pm, \pi_\pm)$ on a constant-$x^-$ slice to the next.

%%%%%%%%%%%%%%%%%%%%%%%%%%%%%%%%%%%%%%%%%
% Figure 1 >>>>>
%%%%%%%%%%%%%%%%%%%%%%%%%%%%%%%%%%%%%%%%%
%%%%%%%%%%%%%%%%%%%%%%%%%%%%%%%%%%%%%%%%%
% Figure  1 <<<<<
%%%%%%%%%%%%%%%%%%%%%%%%%%%%%%%%%%%%%%%%%

Due to the dual-null decomposition, the causal region of a grid is clear, and
there are in-built accuracy checks: the integrability conditions or consistency
conditions $\partial_-\partial_+u=\partial_+\partial_-u$. In order to update a
point N (north), we have two routes from the points E (east) and W (west). 
The sets of equations 
(\ref{eqEinsteinpp})--(\ref{eqEinsteinzz}) [with Eqs. (\ref{eqT++})--(\ref{eqTzz})] 
and (\ref{eq_pip5d})--(\ref{eq_pm5d}) give us the updates  in the $x^+$ direction (from W to N)
and in the $x^-$ direction (from E to N) 
together with the consistency conditions.
Note, however, that there are no equations for $\partial_+\nu_+$, $\partial_-\nu_-$, 
$\partial_\pm\pi_\pm$, and $\partial_\pm p_\pm$, so the consistency on these variables 
will be checked by other methods. More detailed procedures are given in Ref. \cite{ShinkaiHayward}.

As a virtue of the dual-null scheme, we can follow the wormhole throat or
black hole horizons easily. They are both trapping horizons, hypersurfaces
where $\vartheta_+=0$ or $\vartheta_-=0$ \cite{Hayward1994_bhd,Hayward1998_1st}. 
The region between $\vartheta_+=0$ and $\vartheta_-=0$ is recognized as a trapped region if $\theta_+=0$ locates outer ($x^+$ direction), and if 
such a boundary runs null, we can say that a trapped region is a black hole [see Fig.~\ref{fig:2002fig5} (a)]. 

Another benefit is the singular-point excision technique. As we described, the causal region of each
grid point in the dual-null scheme is apparent.  When a grid point is inside a
black hole horizon and near to the singularity, we can exclude that point and
grid points in its future null cone from further numerical computation.

%%%%%%%%%%%%%%%%%%%%%%%%%%%%%%%%%%%%
\subsection{Initial data construction}
%%%%%%%%%%%%%%%%%%%%%%%%%%%%%%%%%%%%
For the preparation of initial data on $\Sigma_0$ ($x^+=x^-=0$), and on $\Sigma_\pm$ $(x^\mp=0, x^\pm>0)$, we 
integrate the set of equations ($\partial_+ $ equations, and $\partial_- $ equations) from the center $\Sigma_0$. 
When we consider static solutions, some additional consistency relations appear. These are
\begin{eqnarray}
\vartheta_++\vartheta_-=0, \\
\nu_++\nu_-=0,\\
\pi_++\pi_-=0,\\
p_++p_-=0,
\end{eqnarray}
which are given from $(\partial_+ +\partial_-)\Omega=0$, $(\partial_++\partial_-)f=0$, $(\partial_++\partial_-)\psi=0$, and $(\partial_++\partial_-)\phi=0$, respectively, together with  
\begin{widetext}
\begin{eqnarray}
e^f\vartheta_+ \nu_+&=&
- \frac{\alpha_{\rm GR}(n-2)(n-3)}{2} \Omega^2 (k+W)+\Lambda + \kappa^2 (V_{\psi}+V_{\phi}) 
\nonumber \\ 
&&
-\frac{\tilde{\alpha}(n-2)(n-5)}{2}\frac{\Omega^3}{A}
\left( k^2+2kW +2W^2 \right)
-\frac{\Omega e^f}{ A}\kappa^2 (\pi_+^2-p_+^2),
\\
e^{f}(\partial_+ +\partial_- +\nu_-)\vartheta_-
&=&\frac{1}{\Omega A}
\left[ -  \frac{\alpha_{\rm GR}(n-2)(n-3)}{2} \Omega^2 (k+W)+\Lambda + \kappa^2 (V_{\psi}+V_{\phi}) \right], 
\nonumber \\ &&
-\frac{\tilde{\alpha}(n-2)(n-5)}{2}\frac{\Omega^3}{A}
\left( k^2+2kW +2W^2 \right)
-\frac{\Omega e^{f}}{ A}\kappa^2 (\pi_-^2-p_-^2), 
\end{eqnarray}
which are given  from 
$(\partial_+ +\partial_-)\vartheta_+=0$ and $(\partial_+ +\partial_-)\vartheta_-=0$, respectively. 

When we consider a static configuration, we have requirements on $\Sigma_0$; 
$\vartheta_+=\vartheta_- =0 $ and $\nu_+=\nu_-=0$.  We also have a constraint on the matter; 
\begin{eqnarray}
 \Omega e^f \kappa^2(\pi_+^2-p_+^2)
&=&\frac{1}{\Omega }
\left[ - \frac{ \alpha_{\rm GR}(n-2)(n-3)}{2} k\Omega^2 +\Lambda + \kappa^2  (V_{\psi}+V_{\phi}) \right]
-\frac{\tilde{\alpha}(n-2)(n-5)}{2} k^2\Omega^3,
\label{constraintatthroat} 
\end{eqnarray}
\end{widetext}
which is derived from $\partial_+\vartheta_\pm=-\partial_-\vartheta_\pm$, and this constraint will be 
concerned when we set $\pi_\pm, p_\pm$ on $\Sigma_0$. 

%%%%%%%%%%%%%%%%%%%%%%%%%%%%%%%%%%%%
\subsection{Transformation from normal metric to dual-null metric}\label{sec2trans}
%%%%%%%%%%%%%%%%%%%%%%%%%%%%%%%%%%%%
In Sec. \ref{section3}, we compare our numerically constructed initial data in a dual-null metric
with the exact solution in normal time-space metric.  Such a transformation is given by the method below. 

Suppose we identify a $(t,r)$ metric 
\begin{eqnarray} 
ds^2&=& -F(t,r) dt^2 +\frac{1}{F(t,r)}dr^2 \\
&=& -F(t,r) (dt^2-dr_\ast^2),
\end{eqnarray}
with a dual-null metric
\begin{eqnarray} 
ds^2&=& -2 e^{f(x^+,x^-)} dx^+ dx^-,  
\end{eqnarray}
where a tortoise coordinate $r_\ast$ is introduced as $\displaystyle \frac{dr}{dr_\ast}=F$. 
By identifying two coordinates as  
\begin{eqnarray} 
x^+&=& \frac{1}{\sqrt{2}}(t+r_\ast),\\
x^-&=& \frac{1}{\sqrt{2}}(t-r_\ast),
\end{eqnarray} 
when we consider a static solution, the derivative of a function 
$G(t,r)$ 
in the $x^+$ direction is expressed as   
\begin{eqnarray} 
\frac{d }{d x^+}G(t,r)&=&\frac{\partial r_\ast}{\partial x^+}\frac{dr}{dr_\ast}\frac{\partial G}{\partial r}=
\frac{F}{\sqrt{2}}\frac{\partial G}{\partial r}.
\end{eqnarray}
Thus, the components in a $(t,r)$ metric can be converted into a $(x^+,x^-)$ metric. 

%%%%%%%%%%%%%%%%%%%%%%%%%%%%%%%%%%%%
\subsection{Misner-Sharp mass}
%%%%%%%%%%%%%%%%%%%%%%%%%%%%%%%%%%%%
In order to evaluate the energy, we apply the Misner-Sharp mass in $n$-dimensional Einstein-GB gravity\cite{MaedaNozawa}, 
\begin{widetext}
\begin{equation}
E_n=\frac{(n-2)A_{n-2}}{2\kappa^2_n\Omega}\left\{-\frac{2\Lambda}{(n-1)(n-2)\Omega^2}
+k+\frac{2e^f}{(n-2)^2}\vartheta_+\vartheta_- 
+\tilde{\alpha} \Omega^2\left[k+\frac{2e^f}{(n-2)^2}\vartheta_+\vartheta_-\right]^2
\right\}, 
\label{misnersharpmassGB}
\end{equation}
\end{widetext}
where $A_{n-2}$ is the volume of the $(n-2)$-dimensional unit constant-curvature space, i.e. 
$A_2=\pi/\Omega^2$, $A_3=4\pi/(3\Omega^3)$, $A_4=\pi^2/\Omega^4$, $A_5=8\pi^2/(15\Omega^5)$  for $k=1$.  
%%%%%%%%%%%%%%%%%%%%%%%%%%%%%%%%%%%%
\subsection{Kretschmann scalar}
%%%%%%%%%%%%%%%%%%%%%%%%%%%%%%%%%%%%
For evaluation of the magnitude of the curvature, we calculate the 
Kretschmann scalar in $n$ dimensions, 
\begin{equation}
{\cal I}^{(n)}=R^{ijkl}R_{ijkl}. 
\end{equation}
${\cal I}^{(n)}$ is written as 
\begin{eqnarray} 
%---------------- n=4
{\cal I}^{(4)}&=&
I_1
+16 I_2
+4 I_3,
\\
%---------------- n=5
{\cal I}^{(5)}
&=&
I_1
+24 I_2
+12 I_3,
\\
%---------------- n=6
{\cal I}^{(6)}&=&
I_1
+32 I_2
+20 I_3
+16 I_4,
\\
%---------------- n=7
{\cal I}^{(7)}&=&
I_1
+40 I_2
+32 I_3
+32 I_4,
\end{eqnarray}
where
%\begin{widetext}
\begin{eqnarray} 
I_1&=&4 e^{2 f} {(\partial_+\partial_- f)}^2,
\\
I_2&=&\frac{e^{2 f}}{r^2} \left\{ 
\bigl[(\partial_- f) (\partial_- r) +(\partial_-\partial_- r)\bigr] \right. \nonumber \\
&&   \times \bigl[(\partial_+ f) (\partial_+ r) +(\partial_+\partial_+ r)\bigl]  
%\nonumber \\ &&  \qquad
 \left.  +(\partial_+\partial_- r)^2 \right\},
\\
I_3&=&\frac{ \left[k+2 e^{2f{}} (\partial_- r) (\partial_+ r)\right]^2}{r{}^4},
\\
I_4&=&\frac{\left[e^{2f}  (\partial_+ r) (\partial_- r)\right]^2}{r^4}.
\end{eqnarray}
%\end{widetext}

%######################################################################%
%######################################################################%
%    SECTION  3 
%######################################################################%
%######################################################################%
%%%%%%%%%%%%%%%%%%%%%%%%%%%%%%%%%%%%%%%%%%%%%%%%%%%%%%%%%%%%%%%%%%%%%%%
\section{Numerical evolutions of a perturbed wormhole}\label{section3}
%%%%%%%%%%%%%%%%%%%%%%%%%%%%%%%%%%%%%%%%%%%%%%%%%%%%%%%%%%%%%%%%%%%%%%%
In this section, we show the evolutions of the Ellis-type wormhole in higher-dimensional space-time 
both in GR and in the Einstein-GB gravity theories. 

In four-dimensional GR, a wormhole is an unstable object. 
If it is perturbed, its throat suffers a bifurcation of horizons and either 
collapses to a black hole, or explodes to form an inflationary universe, 
depending on whether the additional (perturbed) energy is positive or negative, 
respectively (see Fig.~\ref{fig:2002fig5}) \cite{ShinkaiHayward}.

The instability of the Ellis-type wormhole in $n$-dimensional GR is also shown
using a linear perturbation method by us \cite{ToriiShinkaiWH}.  
We showed that the solutions have at least one negative mode, 
which leads to the conclusion that all Ellis-type wormholes are linearly unstable.
The time scale of instability becomes shorter as $n$ becomes larger. 

Therefore, the objectives of this section are to confirm the 
instability of higher-dimensional GR wormholes in the nonlinear regime and to investigate the behavior of Einstein-GB wormholes.

%%%%%%%%%%%%%%%%%%%%%%%%%%%%%%%%%%%%%%%%%
% Figure 2 >>>>
%%%%%%%%%%%%%%%%%%%%%%%%%%%%%%%%%%%%%%%%%
\begin{widetext}
\begin{center}
\begin{figure}[htb]
\includegraphics[width=13cm]{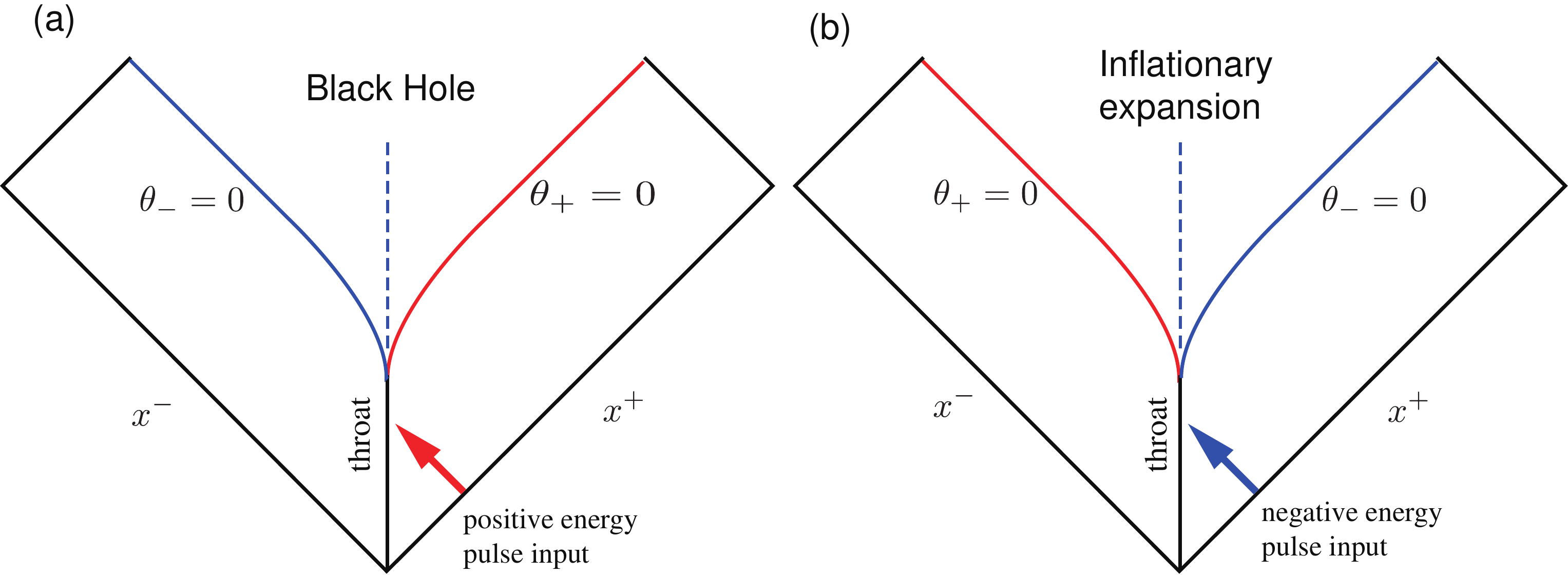}
%\begin{minipage}[b]{20pc}
\caption{\label{fig:2002fig5} Partial Penrose diagrams of the evolved space-time.
Suppose we live in the right-side region and input a pulse to an Ellis wormhole in the 
middle of each diagram. The wormhole throat suffers a bifurcation of horizons and either 
(a) collapses to a black hole, 
or (b) explodes to form an inflationary universe, 
depending on whether the total input
energy is positive or negative, respectively. This basic picture was first given by Ref.~\cite{ShinkaiHayward}, and it holds for higher-dimensional GR as will be shown in Fig.~\ref{fig:evolutionGR}, while in the Einstein-GB gravity slight changes are observed as will be shown in Fig.~\ref{fig:evolutionGB}. 
}
%\end{minipage}
\end{figure}
\end{center}
\end{widetext}
%%%%%%%%%%%%%%%%%%%%%%%%%%%%%%%%%%%%%%%%%
% Figure 2 <<<<
%%%%%%%%%%%%%%%%%%%%%%%%%%%%%%%%%%%%%%%%%
%\clearpage

%%%%%%%%%%%%%%%%%%%%%%%%%%%%%%%%%%%%
\subsection{Wormholes in four-, five- and six-dimensional GR}
%%%%%%%%%%%%%%%%%%%%%%%%%%%%%%%%%%%%
 The solution shown in Ref. \cite{ToriiShinkaiWH} is obtained 
in a spherically symmetric space-time ($k=+1$) with the metric, 
\begin{eqnarray}
%========<Equation>========%
%ds^2 = - f(t,r)e^{-2\delta(t,r)}dt^2 + f(t,r)^{-1} dr^2 + R(t,r)^2 h_{ij}dx^i dx^j,
ds^2 &=& - F(t,r)e^{-2\delta(t,r)}dt^2 + F(t,r)^{-1} dr^2 \nonumber \\
&&+ R(t,r)^2 \gamma_{ij}dz^i dz^j,
\label{metric_1}
\end{eqnarray}
with a massless ghost scalar field ($V_\phi=0$). 
In order to construct a static wormhole solution, the metric function is restricted 
as $F=F(r)$, $R=R(r)$, $\phi=\phi(r)$, and $\delta=0$. 
By locating the throat of the wormhole at $r=0$, 
and imposing the reflection symmetry at the throat, 
the solution of the field equations is obtained as 
\begin{eqnarray}%========<Equation>========%
f\equiv 1, \qquad
R' = \sqrt{1-\Big(\frac{a_0}{R}\Big)^{2(n-3)}}, \nonumber \\
%~~~{\color{green} \biggl[R'' = \frac{n-3}{R^{2n-5}}  \biggr]}
\phi= \frac{\sqrt{(n-2)(n-3)}}{\kappa} a_0^{n-3}  \int\frac{1}{R(r)^{n-2}}dr, 
\label{solToriiShinkai}
\end{eqnarray}
where $a_0$ is the radius of the throat, i.e. $R(0)=a_0$, 
and a prime denotes a derivative with respect to $r$. 
We used this solution for confirmation of our numerical solution, 
using the method described in Sec. \ref{sec2trans}.

In order to construct the initial static data on $\Sigma_\pm$, 
we integrate $x^+$ equations [(\ref{eqEinsteinpp}), (\ref{eqEinsteinpm}) as $\partial_+\vartheta_-$, (\ref{eqEinsteinzz}), (\ref{eq_pip5d}), and (\ref{eq_pp5d})] and 
$x^-$ equations [(\ref{eqEinsteinmm}), (\ref{eqEinsteinpm}), (\ref{eqEinsteinzz}) as $\partial_- \nu_+$, 
(\ref{eq_pim5d}), and (\ref{eq_pm5d})] with the boundary values at the throat, 
\begin{equation}
\Omega=\frac{1}{a_0}, \quad
\vartheta_\pm=\nu_\pm=f=0, \quad
\phi=\phi_0, 
\end{equation}
where $\phi_0$ is given by Eq.~(\ref{solToriiShinkai}), 
and we set $p_\pm (<0)$ from Eq.~(\ref{constraintatthroat}). 
 
\begin{widetext}
\begin{center}
%%%%%%%%%%%%%%%%%%%%%%%%%%%%%%%%%%%%%%%%%
% Figure 3 >>>>
%%%%%%%%%%%%%%%%%%%%%%%%%%%%%%%%%%%%%%%%%
\begin{figure}[htb]
\includegraphics[keepaspectratio=true,width=6.25cm]{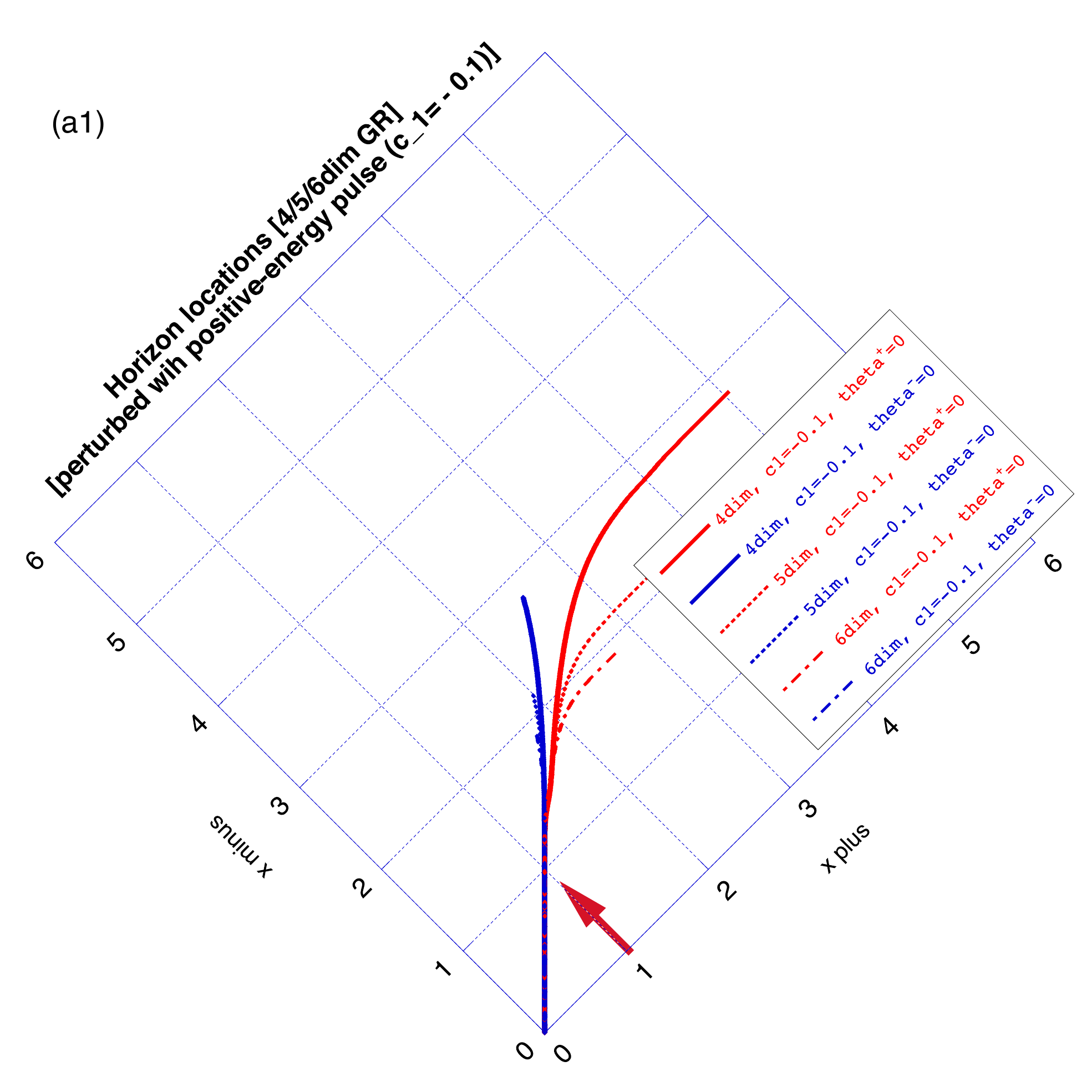}~~~
\includegraphics[keepaspectratio=true,width=6.25cm]{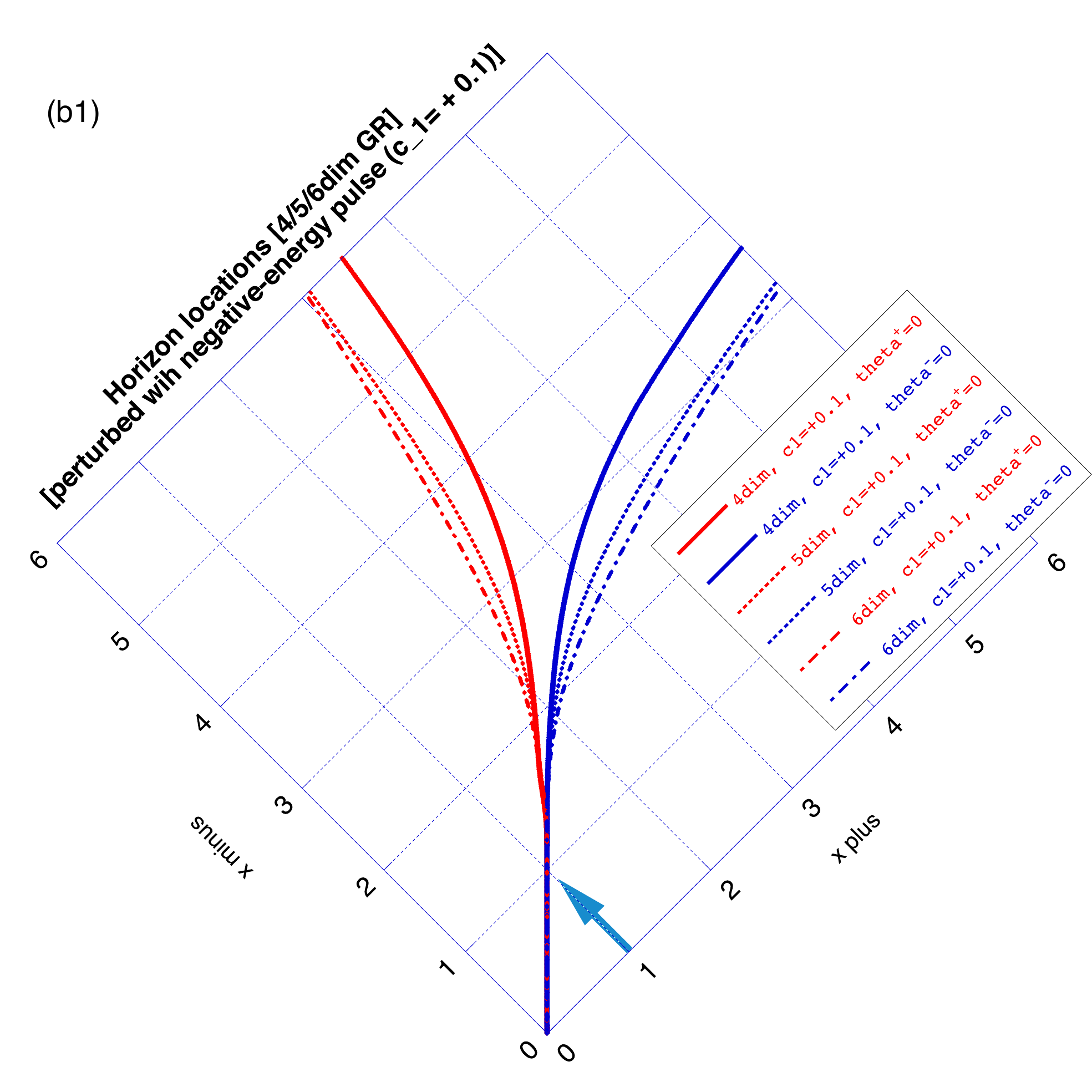}\\~\\
\includegraphics[keepaspectratio=true,width=6.25cm]{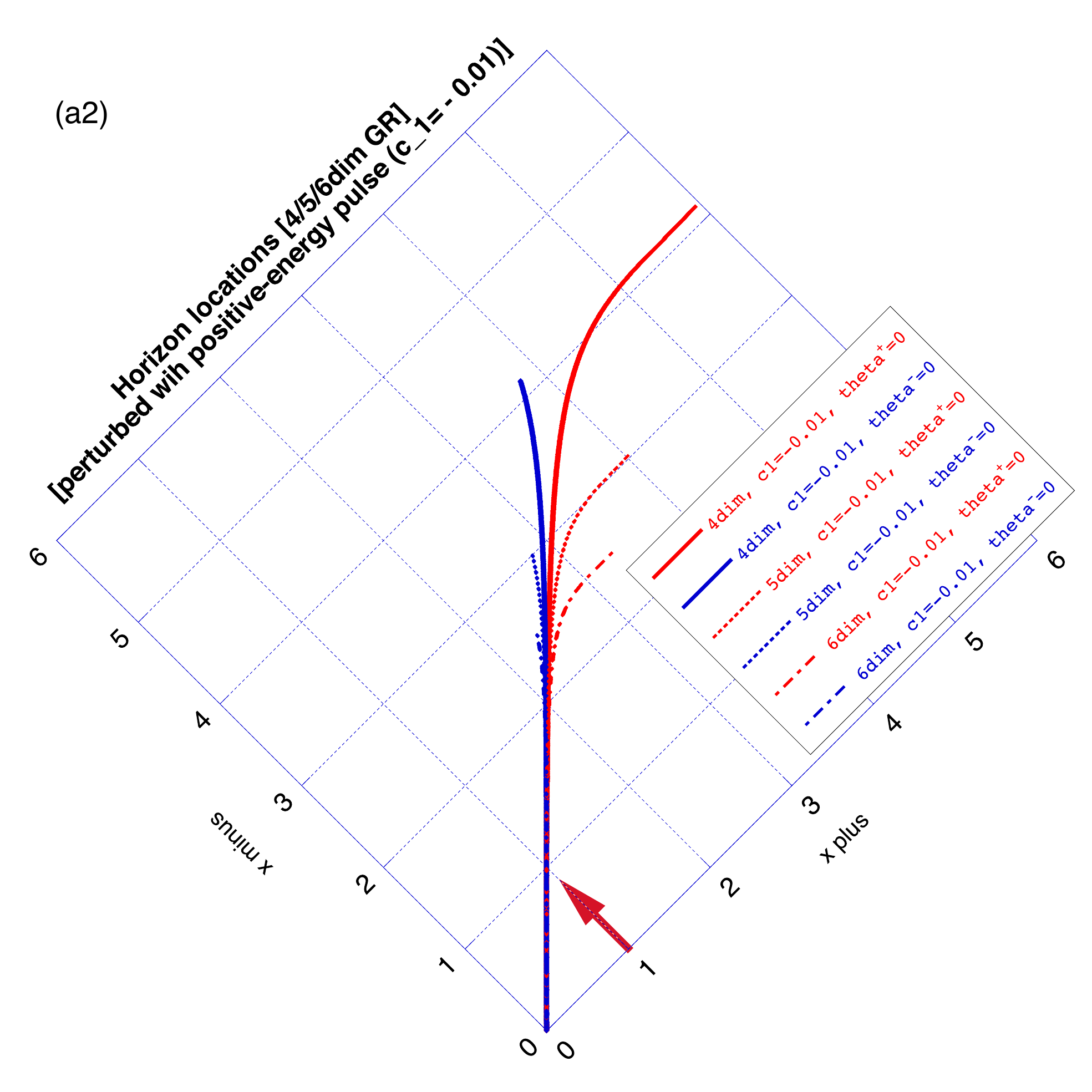}~~~
\includegraphics[keepaspectratio=true,width=6.25cm]{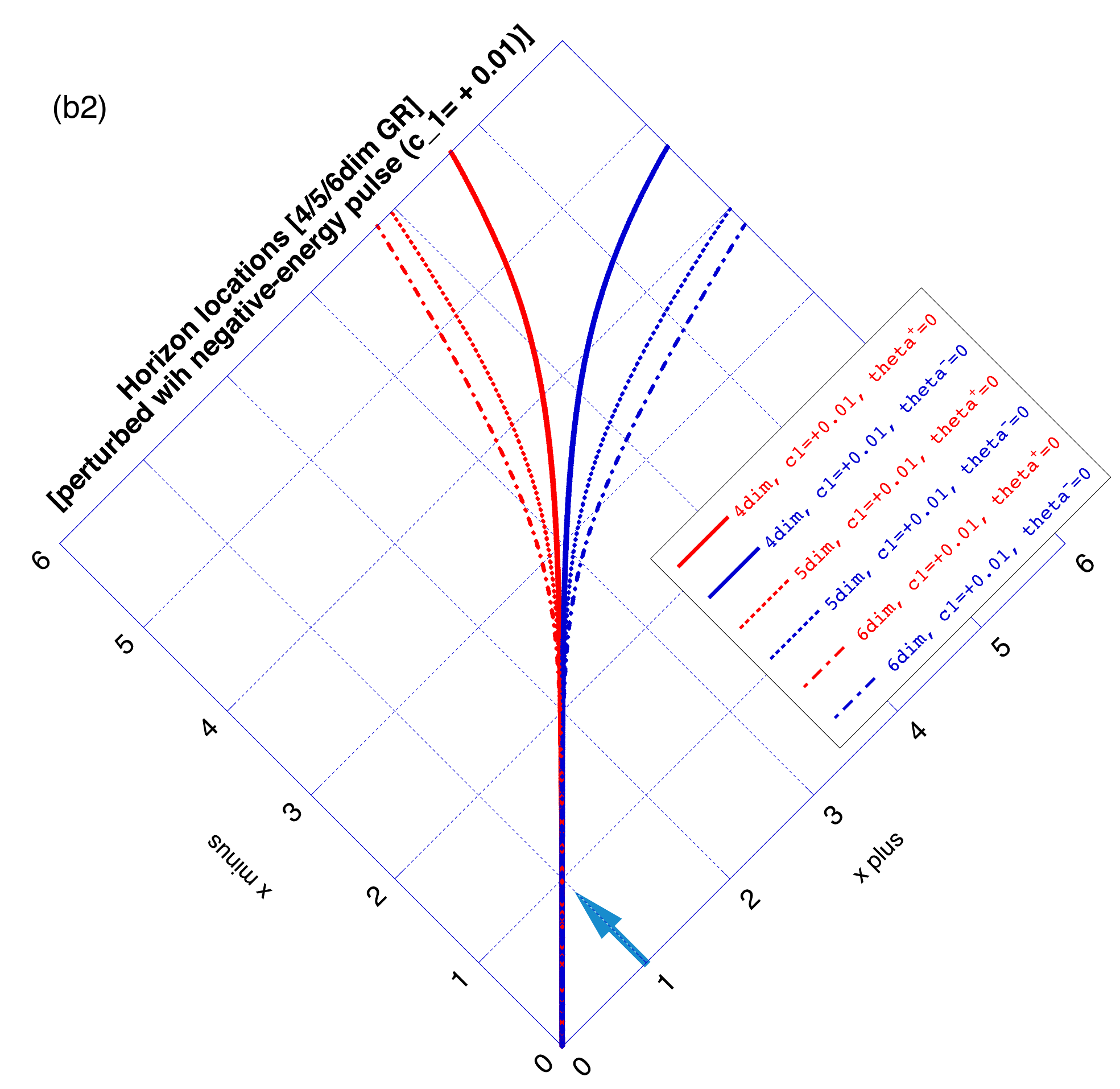}\\
\caption{\label{fig:evolutionGR}
Evolutions of a perturbed wormhole in four-, five-, and six-dimensional GR 
($\alpha_{\mbox{\footnotesize GB}}=0$). 
Locations of the horizons [where the expansions are $\vartheta_+=0$ (red lines) 
and $\vartheta_-=0$ (blue lines)] are plotted as a function of $(x^+, x^-)$. 
Figures (a1) and (a2) are the results of the injection of a 
positive-energy scalar pulse which hits the throat at $x^+=x^-=1$, while Figures 
(b1) and (b2) are those of a negative-energy pulse. Arrows indicate the trajectories of pulses. 
The pulse parameters are $c_1=-0.1$ for (a1), $c_1=0.1$ for (b1), $c_1=-0.01$ for (a2) and $c_1= 0.01$ for (b2).
We also set  $c_2=3$ and $c_3=1$, which means that the pulse hits the wormhole throat at $x^+=x^-=1$. 
The throat begins turning into a black hole
 if we input positive-energy scalar flux (left panels), while the throat 
expands if we input negative-energy scalar flux (left panels). 
This is what we expected from Fig.~\ref{fig:2002fig5}.  
We also see that the bifurcation of the throat appears earlier for higher dimensions, which suggests larger instability.  The figures should be symmetric, but the large curvature stops numerical evolution just after a black hole is formed, 
so that the plots in the left panels are terminated in the middle of $x^-$. 
}
\end{figure}
%%%%%%%%%%%%%%%%%%%%%%%%%%%%%%%%%%%%%%%%%
% Figure 3 <<<<
%%%%%%%%%%%%%%%%%%%%%%%%%%%%%%%%%%%%%%%%%
\end{center}
\end{widetext}
%\clearpage

We find that numerical truncation error can quite easily destroy the static configuration, 
but  it can be controlled with finer resolution.  All the results below are shown 
after we have confirmed that the static solution of the wormhole is maintained 
during the evolution (in $x^-$ direction) in the range of discussion. 

We add a perturbation to the static wormhole in the form of Gaussian
pulse, input from the right-hand universe. 
The perturbation is placed as scalar-field momentum on the initial data $\Sigma_+$ as a form 
\begin{equation}
\delta p_+= c_1 \exp \big[-c_2(x^+-c_3)^2\bigr], \label{pulseeq}
\end{equation}
for the ghost scalar field where $c_1, c_2, c_3$ are parameters, 
or
\begin{equation}
\delta \pi_+= c_1 \exp \big[-c_2(x^+-c_3)^2\bigr], \label{pulseeq2}
\end{equation}
for the normal scalar field. 
The static wormhole solution consists from the ghost field, and its total energy is zero.  In this model, 
positive (negative) $c_1$ in the ghost field (\ref{pulseeq}) 
indicates the addition of positive (negative) energy to the system, while 
$c_1\neq 0$ in the normal field (\ref{pulseeq2}) indicates the addition of positive energy to the system. 
After we set this perturbation form, 
we re-solve the other variables on $\Sigma_+$;  
i.e., our perturbed initial data are all solutions of the system, and we can also 
add a perturbation beyond the linear level.

Figure~\ref{fig:evolutionGR} shows the results of four-, five-, and six- dimensional wormhole
solutions with the above perturbations.  
The plots show the trajectories of the locations of vanishing expansions  
$\vartheta_\pm=0$ in the $(x^+,x^-)$ plane.  
We see that the wormhole throat is initially located where  $\vartheta_+=\vartheta_-=0$, but 
after a small pulse hits it, the throat (or horizon) splits into two horizons ($\vartheta_+=0$
and $\vartheta_-=0$),  and they propagate in opposite directions, depending on the 
signature of the energy of the pulse.

If the location of $\vartheta_+=0$ is farther out (in the $x^+$ direction) 
than that of $\vartheta_-=0$, then the region 
between $\vartheta_-=0$ and $\vartheta_+=0$ is said to be {\it trapped}. 
If such a trapped surface runs null, then the region is judged to be a black hole. 
On the contrary, if  $\vartheta_-=0$ is farther out  (in the $x^+$ direction), then 
the region between $\vartheta_+=0$ and $\vartheta_-=0$ can be judged as an expanding throat. 
These two differences are confirmed also by calculating the circumference radius (see Fig.~\ref{fig:evolutionGBthroat}, later).

The throat begins shrinking and turns into a black hole if we inject a 
positive-energy scalar flux (left panels in Figure \ref{fig:evolutionGR}), 
while the throat begins expanding if we input a negative-energy scalar flux (right panels). 
This fundamental feature is the same with those already reported in Ref.~\cite{ShinkaiHayward}, and the fact that higher-dimensional cases show earlier bifurcation matches with the predicted behavior from the linear perturbation analysis in Ref.~\cite{ToriiShinkaiWH}.

%%%%%%%%%%%%%%%%%%%%%%%%%%%%%%%%%%%%
\subsection{Wormholes in Einstein-GB gravity}
%%%%%%%%%%%%%%%%%%%%%%%%%%%%%%%%%%%%
We also evolved the perturbed wormhole initial data with the GB terms ($\alpha_{\rm GB}\neq 0$)
and studied their effects on the evolutions.  
We first prepared the static Ellis-type wormhole solution by solving equations on $\Sigma_\pm$ numerically. 
We checked that the solution in $n=4$ Einstein-GB 
gravity is identical with that in GR. 

We then confirmed that the solution is static by evolving it without perturbation. 
We actually found that the evolutions with large $|\alpha_{\rm GB}|$ are quite unstable numerically, and it is hard to keep its static configurations long enough.
Therefore we can present the results only for small-$|\alpha_{\rm GB}|$ cases, for those we confirmed the static configuration is maintained for the range of discussion. 

Figure~\ref{fig:evolutionGB} shows the cases of $n=5$ and 6 Einstein-GB gravity with $\alpha_{\rm GB}=+0.001$. 
The lines show the locations of horizons ($\vartheta_\pm=0$). 
We change the amplitude of the perturbation, $c_1$ in Eq. (\ref{pulseeq}), and find that for large $c_1$,  
the throat turns into a black hole, while for small $c_1$, the throat begins expanding.  
This statement will be clarified in Fig.~\ref{fig:evolutionGBthroat}.

Figure~\ref{fig:evolutionGBthroat} shows the evolution behavior of the circumference radius of the throat. 
We plotted for the cases of Fig.~\ref{fig:evolutionGB}. 
If the amplitude of the perturbation, $c_1$, is above a particular value, then the throat begins shrinking, which indicates the formation of a black hole. 
The critical value of the parameter exists at $E\sim 1.0$ for $n=5$ and at $E\sim 2.5$ for $n=6$ in terms of the Misner-Sharp mass (\ref{misnersharpmassGB}), which means the threshold is larger for $n=6$.
Since the energy of the injected pulse is always positive, the existence of the critical positive value for forming a black hole suggests that introducing the GB terms with $\alpha_{\rm GB}>0$ turns out to have a sort of ``negative" energy. The larger threshold for forming a black hole in higher dimensions also indicates that such effects become stronger in higher dimensions. 

For quantitative comparisons, we prepared Table~\ref{table1}, 
in which we list how the initial (positive-energy) perturbation, $\Delta E$, results in a black hole (if it is formed). 
We evaluated the Misner-Sharp mass (\ref{misnersharpmassGB})  at the end of the grid, $x^+=5$, and measured the horizon coordinate, $x^-_H$
where the outgoing trapping horizon, $\theta_+=0$ propagates at null. 
We see that in higher $n$,  $x^-_H$ is smaller, 
which indicates the early formation of a black hole due to the large instability. 
Interestingly, the final mass of the black hole, $E_f$,  depends only on the dimension $n$ and $\alpha_{\rm GB}$, and does not depend on the injected energy, $\Delta E$.  The black hole mass, $E_f$, is supposed to be a critically formed minimum mass of the black hole, and such an existence of the minimum mass (or threshold) was the same with those in the four-dimensional GR cases \cite{ShinkaiHayward}. 
This threshold is larger for large $\alpha_{\rm GB}$. 
The listed cases are fixed by the amplitude, $c_1$, of the injected perturbation, but if we check the ratio $E_f/E_i$,  then we see that the final mass of the black hole becomes smaller when $\alpha_{\rm GB}$ is larger.
Both suggest that the GB terms work for avoiding the appearance of a black hole (or singularity). 

\begin{widetext}
\begin{center}
%%%%%%%%%%%%%%%%%%%%%%%%%%%%%%%%%%%%%%%%%
% Figure 4 >>>>
%%%%%%%%%%%%%%%%%%%%%%%%%%%%%%%%%%%%%%%%%
\begin{figure}[ht]
\includegraphics[keepaspectratio=true,width=7cm]{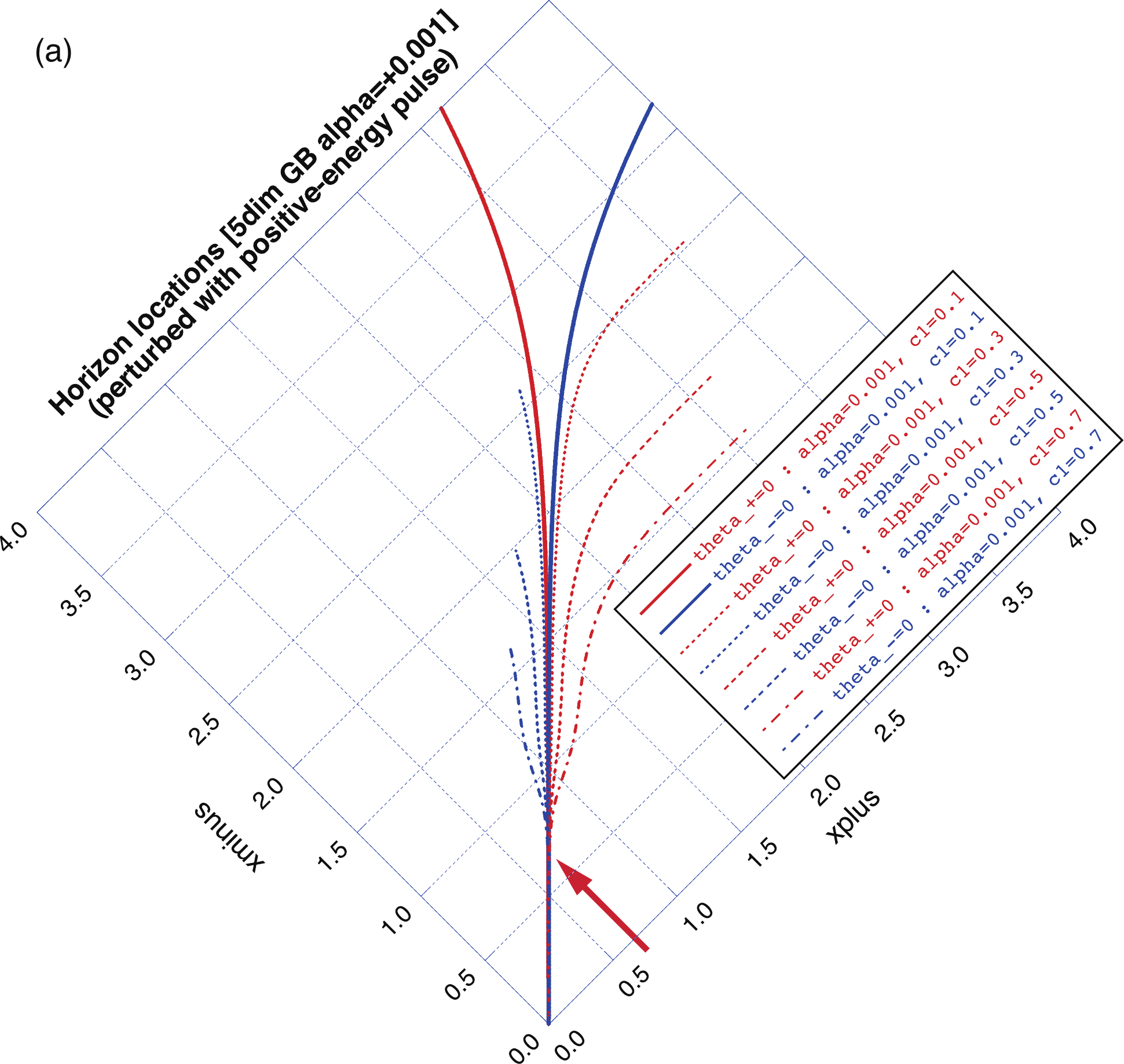}~~~
\includegraphics[keepaspectratio=true,width=7cm]{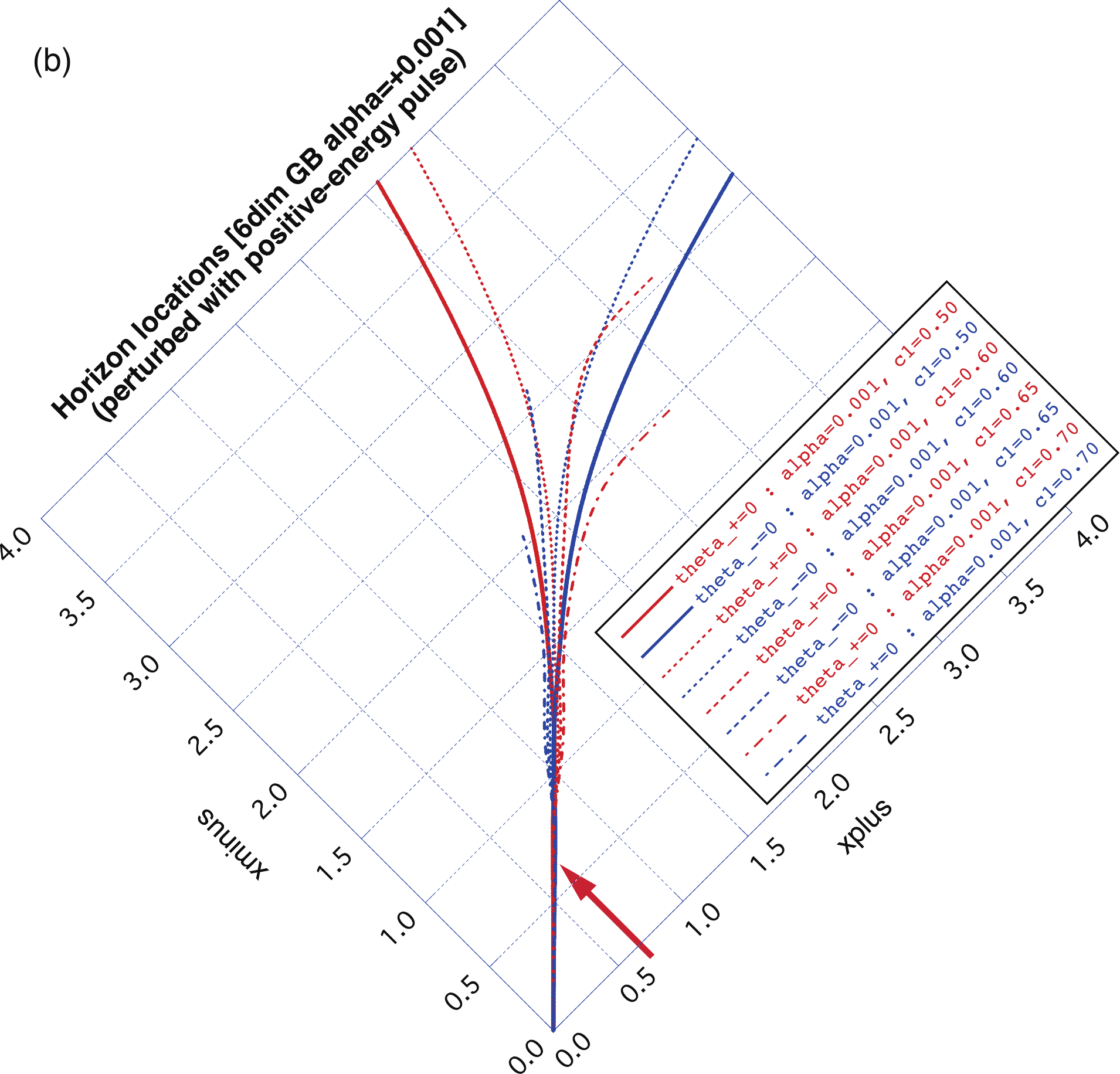}
\caption{\label{fig:evolutionGB}
Evolutions of a perturbed wormhole in Einstein-GB gravity with $\alpha_{\rm GB}=+0.001$.  
The left and right panels show the cases of five- and six-dimensional space-time, respectively.  
Locations of the horizons [where the expansions are $\vartheta_+=0$ (red lines) 
and $\vartheta_-=0$ (blue lines)] are plotted as a function of $(x^+,\: x^-)$
for several amplitudes of the perturbation (\ref{pulseeq}) with $c_1=0.3,\; 0.5,\;0.7$ for $n=5$, and 
$c_1=0.5,\; 0.6,\; 0.65, \;0.7$ for $n=6$.
The other parameters of the injections are $c_2=16$ and $c_3=0.7$. 
Arrows indicate the trajectories of pulses.
We see that for large $c_1$, the throat turns into a black hole, 
while for small $c_1$, the throat begins expanding, which is different from GR cases.  
}
\end{figure}
%%%%%%%%%%%%%%%%%%%%%%%%%%%%%%%%%%%%%%%%%
% Figure 4 <<<<
%%%%%%%%%%%%%%%%%%%%%%%%%%%%%%%%%%%%%%%%%
\end{center}

%\clearpage

%\end{widetext}
%\begin{widetext}
\begin{center}
%%%%%%%%%%%%%%%%%%%%%%%%%%%%%%%%%%%%%%%%%
% Figure 5 >>>>
%%%%%%%%%%%%%%%%%%%%%%%%%%%%%%%%%%%%%%%%%
\begin{figure}[ht]
~\vspace{-2cm}\\
\includegraphics[keepaspectratio=true,width=8cm]{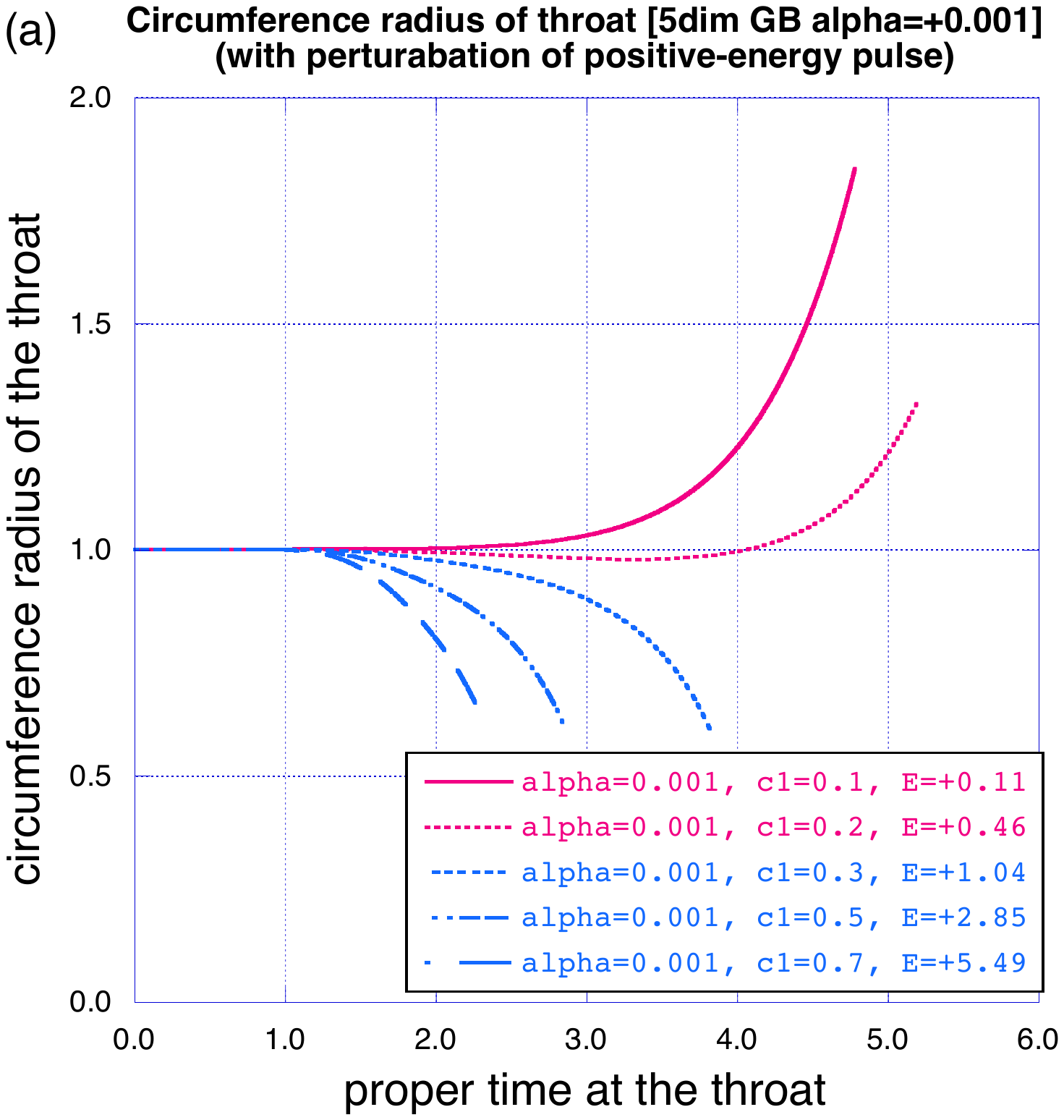} ~ 
\includegraphics[keepaspectratio=true,width=8cm]{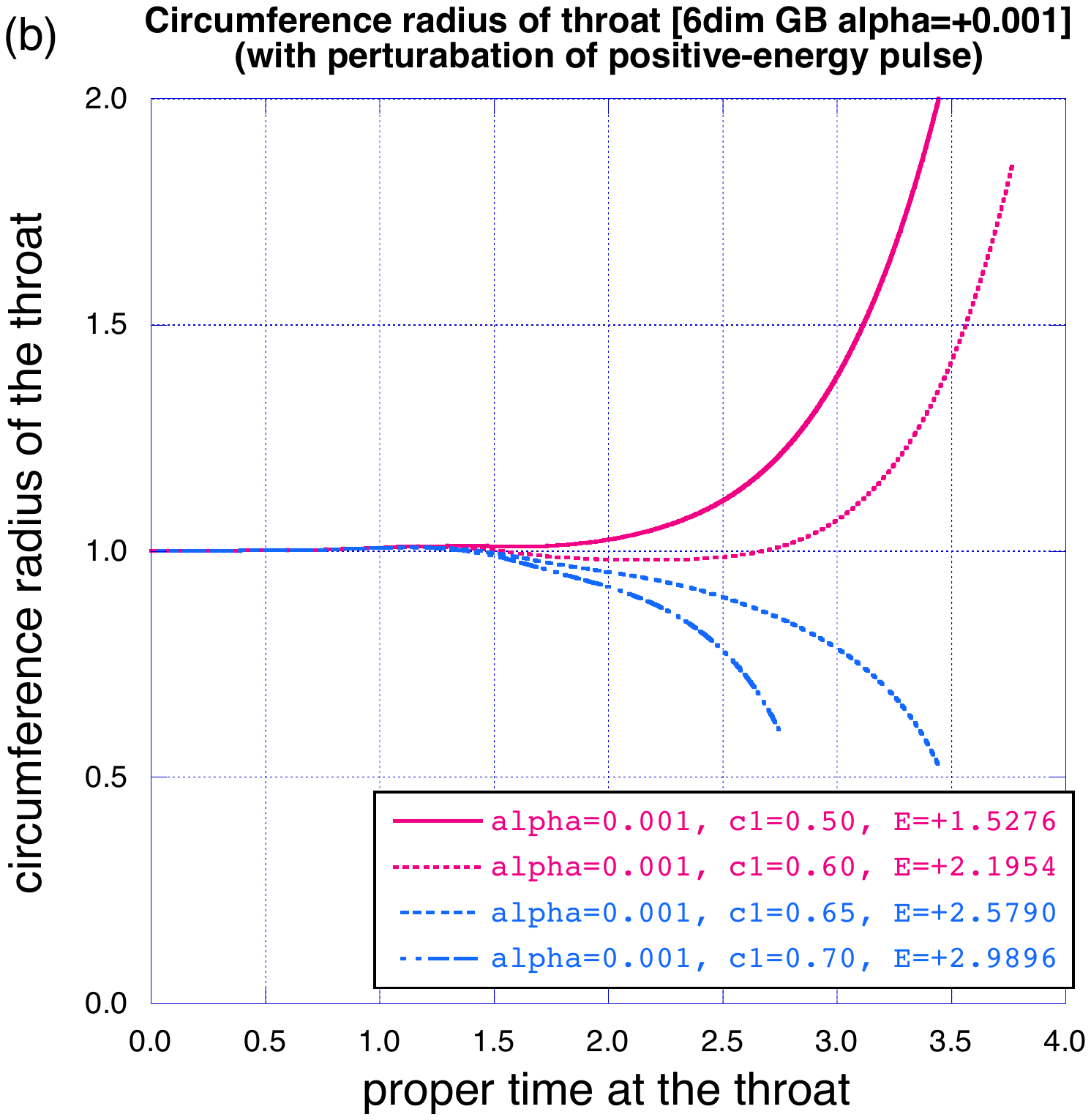}~\\
~\vspace{-3cm}
\caption{\label{fig:evolutionGBthroat}
The behavior of the circumference radius of the throat for the cases of Fig.~\ref{fig:evolutionGB}. 
Panel (a) shows the cases of five-dimensional space-time, while (b) gives the cases of six-dimensional space-time.
We see that if the amplitude of the perturbation, $c_1$, is above a particular value, the throat begin shrinking, which indicates the formation of a black hole. This critical value is expressed with the Misner-Sharp mass   (\ref{misnersharpmassGB}), and we find that the magnitude is larger for $n=6$.
}
\end{figure}
%%%%%%%%%%%%%%%%%%%%%%%%%%%%%%%%%%%%%%%%%
% Figure 5 <<<<
%%%%%%%%%%%%%%%%%%%%%%%%%%%%%%%%%%%%%%%%%
\end{center}
\end{widetext}

%%%%%%%%%%%%%%%%%%%%%%%%%%%%%%%%%%%%%%%%%
% Figure 6 >>>>
%%%%%%%%%%%%%%%%%%%%%%%%%%%%%%%%%%%%%%%%%
\begin{figure}[hbt]
\includegraphics[keepaspectratio=true,width=7cm]{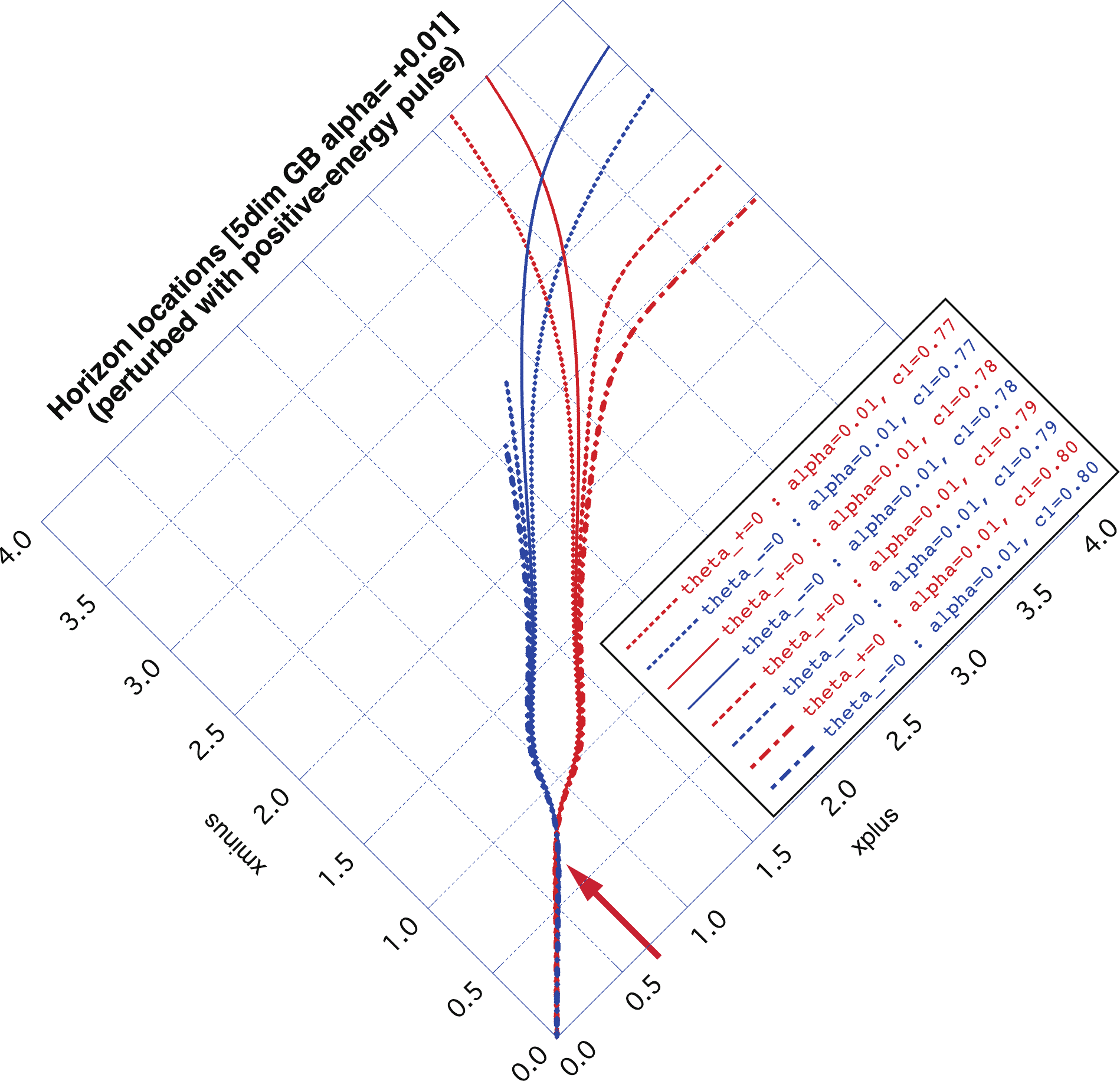}
\caption{The evolutions of the wormhole in the Einstein-GB theory (five-dimensional, $\alpha_{\rm GB}=+0.01$). 
The locations of the horizons are plotted. 
When the amplitude of the perturbation, $c_1$, is close to the critical value for the fate of the wormhole
(either to expansion or to a black hole), a temporal trapped region with a constant radius appears. 
This behavior also suggests that the existence of such a trapped surface
is not a necessary condition for forming a black hole in the Einstein-GB theory. 
}
\label{fig:GBthroatcritical}
\end{figure}
%%%%%%%%%%%%%%%%%%%%%%%%%%%%%%%%%%%%%%%%%
% Figure 6 <<<<
%%%%%%%%%%%%%%%%%%%%%%%%%%%%%%%%%%%%%%%%%

%\clearpage

%%%%%%%%%%%%%%%%%%%%%%%%%
%  TABLE >>>>
%%%%%%%%%%%%%%%%%%%%%%%%%
%>>>>>>>>>>>>>>>>>>>>>>>>>>>>>>>>>>>>>>> Table.\ref{table1}
\begin{table}[bt]
\caption{\label{table1}
Injected perturbation and the final black hole structure (when it is formed). 
Initial Misner-Sharp energy $\Delta E$, Eq. (\ref{misnersharpmassGB}), is the additional energy due to the injected part. 
The amplitude $c_1$ in Eq. (\ref{pulseeq2}) is listed, while we set $c_2=16$ and $c_3=0.7$ for all cases. 
The total energies, $E_i$ and $E_f$, are evaluated at $x^+=5$,  and $E_f$ is regarded as the mass of the black hole  (when it is formed).
The horizon coordinate $x^-_H$ is evaluated where the $\vartheta_+=0$ trapping horizon becomes null.  
}
\begin{tabular}{cc||ccc|c|ccl}
\hline
$n$ & $\alpha_{\rm GB}$ &\multicolumn{3}{c|}{injected field} & Initial& \multicolumn{2}{c}{final BH}&
\\
\hline
 & &Field& $c_1$ & $\Delta E/a_0$& $E_i/a_0$  & $E_f/a_0$ & $x^-_H/a_0$ & 
\\
\hline
\hline
4& $0~~~~~$&$\pi_+$&$+0.25$  &  $+0.03$ & $0.88$ & $3.14$ &  $2.94$  & \\ %w4003R OK 
4& $0~~~~~$&$\pi_+$&$+0.50$  &  $+0.10$ & $0.95$ & $3.14$ &  $2.09$  & \\ %w4005R OK
\hline
5& $0~~~~~$&$\pi_+$&$+0.25$  &  $+0.15$ & $0.52$ & $6.28$ &  $2.26$  & \\ %w5002R OK
5& $0~~~~~$&$\pi_+$&$+0.50$  &  $+0.61$ & $0.97$ & $6.28$ &  $1.66$  &  \\ %w5005R OK
\hline
6& $0~~~~~$&$\pi_+$&$+0.25$  &  $+0.38$ & $0.50$ & $9.87$ &  $1.92$  & \\ %w6003R OK
6& $0~~~~~$&$\pi_+$&$+0.50$  &  $+1.50$ & $1.63$ & $9.87$ &  $1.46$  & \\ %w6005R OK
\hline
\hline
5& $0.001$&$\pi_+$&$+0.25$  &  $+0.15$  & $0.53$ & $6.30$&  $2.67$   & \\ %w5d03R OK
5& $0.001$&$\pi_+$&$+0.50$  &  $+0.61$  & $0.98$ & $6.30$&  $1.72$   & \\ %w5d05R OK
5& $0.001$&$\pi_+$&$+1.00$  &  $+2.23$  & $2.61$ & $6.30$&  $0.98$   & \\ %w5d0tR OK
\hline
5& $0.01~$&$\pi_+$&$+0.50$  &  $+0.59$  & $1.02$ &  &   & noBH\\ %w5c050 OK
5& $0.01~$&$\pi_+$&$+0.75$  &  $+1.31$  & $1.74$ & $6.41$&  $1.92$  & \\ %w5c07R OK
5& $0.01~$&$\pi_+$&$+1.00$  &  $+2.21$  & $2.65$ & $6.41$&  $1.19$  & \\ %w5c0tR OK
\hline
6& $0.001$&$\pi_+$&$+0.50$  &  $+1.53$  & $1.46$&  &    & noBH \\ %w6d050   both LR
6& $0.001$&$\pi_+$&$+0.75$  &  $+3.42$  &$3.36$& $9.93$ &  $1.34$  & \\ %w6d07R OK
6& $0.001$&$\pi_+$&$+1.00$  &  $+6.07$  &$6.60$& $9.93$ &  $1.00$  & \\ %w6d0tR OK
\hline 
6& $0.01~$&$\pi_+$&$+1.00$  &  $+6.90$ & $5.00$ &  &  & noBH\\ %w6c0t0  both LR
6& $0.01~$&$\pi_+$&$+1.50$  &  $+8.78$ & $8.15$ &  &  & noBH\\ %w6c0f0  both LR
\hline
\end{tabular}
\end{table}
%<<<<<<<<<<<<<<<<<<<<<<<<<<<<<<<<<<<<<<<       \ref{table1}
%%%%%%%%%%%%%%%%%%%%%%%%%
%  TABLE <<<<
%%%%%%%%%%%%%%%%%%%%%%%%%

One more interesting finding is the critical case.  
When we tune the perturbation amplitude  $c_1$  close to the critical value, as we show in 
Fig.~\ref{fig:GBthroatcritical}, we find that
the throat (double trapping horizon, $\vartheta_\pm=0$) bifurcates to two trapping horizons ($\vartheta_+=0$ and $\vartheta_-=0$), and 
they remain at a {\it quasi}-constant radius, and shortly after that they propagate outward. 
That is, the wormhole first changes to a 
{\it temporal} trapped region, 
and then decides its fate towards either a black hole or to an expanding throat. 
Actually, the circumference radius of the throat in this critical case takes the value between the red lines and blue lines in Fig. \ref{fig:evolutionGBthroat}; i.e., it remains almost constant but oscillates slightly when it is forming a {\it temporal}  trapped region. 
%Here, we use the word  {\it temporal}  as its throat's circumference radius (Fig.~\ref{fig:evolutionGBthroat}) is kept almost at the constant before it goes to its final object. 
Since the final two objects are totally different and there is no static configuration between them, we guess that this is the first-order transition.  

This critical behavior also suggests us that the existence of such a trapped region 
is not a necessary condition for forming a black hole in this model.  
We do not know if such an observation is general 
in the presence of the GB terms, or if this is only due to the effect of the ghost field. 
We, however, note that, in Einstein-GB gravity, a couple of examples of the differences (from GR) in causality and energy conditions have been reported (e.g. Refs. \cite{MaedaNozawa, Izumi2014}).
Therefore, this new finding might not be surprising.

%######################################################################%
%######################################################################%
%    SECTION  4 
%######################################################################%
%######################################################################%
\section{Numerical evolutions of the collision of scalar pulses}\label{section4}
In this section, we show our results of the collision of massless scalar pulses 
in plane-symmetric space-time.  
There are several exact solutions of the colliding plane waves, which produce curvature singularity after their collisions (see,  e.g.,  Ref. \cite{GriffithsBook} and references therein). 
We prepare a similar situation in our code and examine such a strong curvature effect in higher-dimensional GR and in Einstein-GB gravity. 
We first note that in the construction of exact solutions, the wave fronts are assumed to be a step function, while in our simulations the wave fronts are a continuous function. %Since we prepare the initial data by solving all the field equations, we can present the effects of both perturbative and nonlinear regimes of the pulses. 

We put a perturbed normal scalar field ($\psi$) in the flat background on the initial surfaces and evolve it. 
The space-time is assumed to be plane symmetric  ($k=0$ in Sec. 2), and we do not consider the ghost scalar field ($\phi$) in this section. 
The initial scalar field is set as $\psi=0$ and has momentum 
\begin{eqnarray}%
\left\{%
\begin{array}{l}
\pi_+=a \exp (-b (x^+/\sqrt{2}-c)^2)\\
\pi_-=0
\end{array} \right. 
&~&\mbox{on~$\Sigma_+$}
\label{SPpars}
\\
\left\{%
\begin{array}{l}%
\pi_+=0\\
\pi_-=a \exp (-b (x^-/\sqrt{2}-c)^2) 
\end{array} \right. 
&~&\mbox{on~$\Sigma_-$}
\label{SPpars2}
\end{eqnarray}

where $a, b, c$ are parameters. 

\subsection{Evolutions in GR}
The two typical evolutions are shown in Fig.~\ref{fig:SPcollision}.  We plot the behaviors of the scalar field and the 
Kretschmann scalar, ${\cal I}^{(5)}$, for five-dimensional GR. We set $a=0.2$ and 0.4, and $b=10$, $c=2$ for these plots. 
For small pulses [Fig.~\ref{fig:SPcollision}(a)],  we see that two pulses just pass through each other and the curvature ${\cal I}^{(5)}$ turns back to the flat again. 
On the contrary, for large pulses [Fig.~\ref{fig:SPcollision}(b)], the nonlinear curvature evolution appears after the collision of pulses.  The latter behavior is similar to the exact solutions of the plane-wave collision (see, e.g., figures in Ref. \cite{MatznerTipler1984}). 
We actually find that in all blow-up regions, both expansions are $\vartheta_\pm <0$ (Fig.~\ref{fig:thetaplusminus}). In four-dimensional plane-symmetric space-time, if the curvature blows up, then it means the appearance of a naked singularity, since there is no chance to form a horizon. 
However, in higher dimensions, we expect such a blow-up will be hidden in a horizon as the expansions suggest.

%\clearpage
\begin{widetext}
%%%%%%%%%%%%%%%%%%%%%%%%%%%%%%%%%%%%%%%%%
% Figure 7 >>>>
%%%%%%%%%%%%%%%%%%%%%%%%%%%%%%%%%%%%%%%%%
\begin{center}
\begin{figure}[tbh]
%\begin{tabular}{p{16cm}}
\includegraphics[keepaspectratio=true,width=14cm]{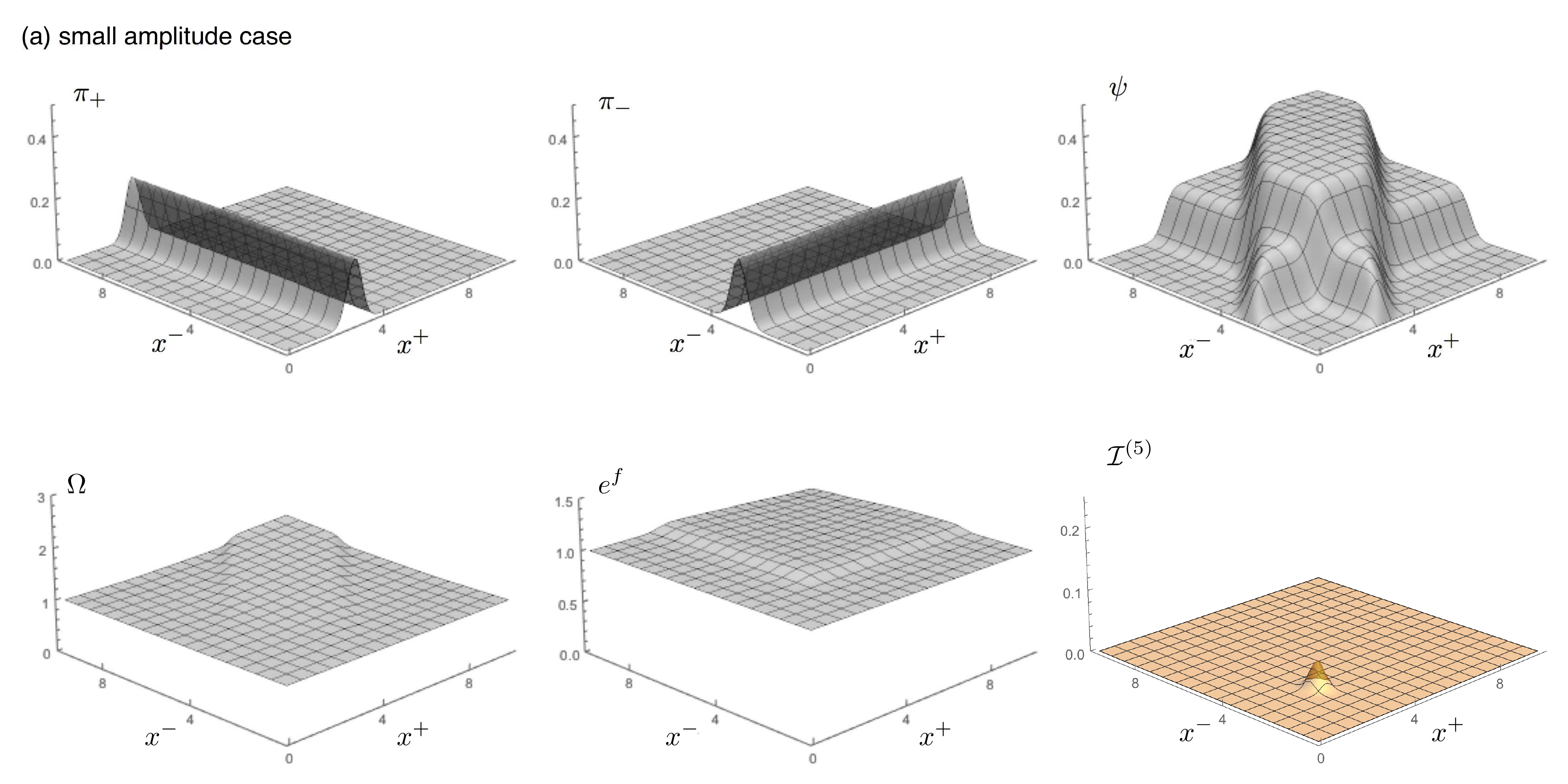}\\
\includegraphics[keepaspectratio=true,width=14cm]{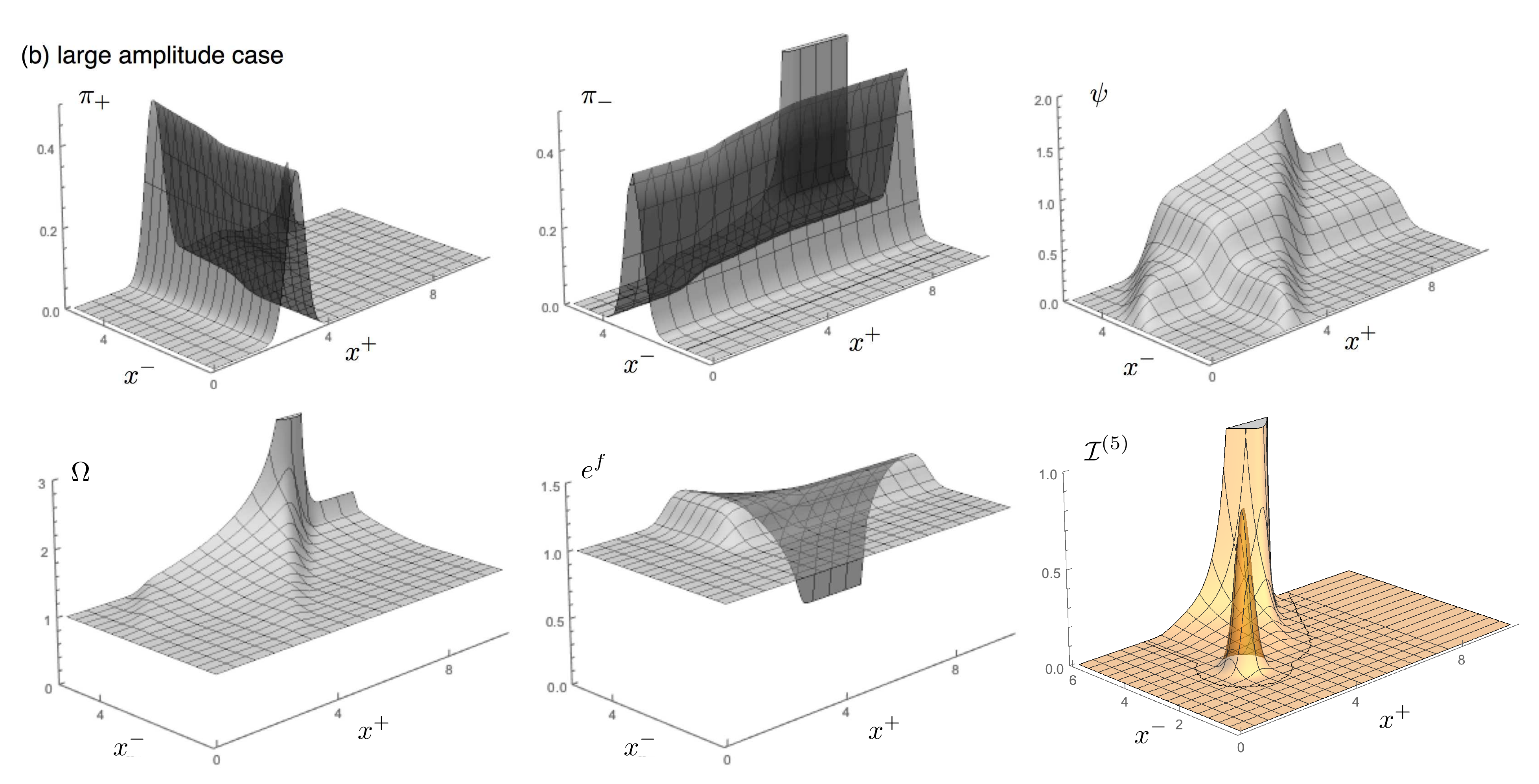}
%\end{tabular}
\caption{\label{fig:SPcollision} 
Evolutions of colliding two scalar pulses in five-dimensional GR: (a) the small-amplitude case [$a=0.2$ in Eqs. (\ref{SPpars}) and (\ref{SPpars2})], and  (b) the large-amplitude case ($a=0.4$). The scalar momentum $\pi_\pm$, scalar field $\psi$,  the conformal factor $\Omega$, metric function $e^{f}$, and the Kretschmann scalar ${\cal I}^{(5)}$ are plotted in the $(x^+,\:x^-)$ coordinates. Initial data were set at both $\Sigma_-\;(x^+=0,\:x^->0)$ and  $\Sigma_+\;(x^+>0,\:x^-=0)$ and evolved. For small pulses, we see that they just cross, and  space-time turns back toward flat again, while for large pulses,  we see that nonlinear curvature evolution appears after the collision of pulses.  The latter behavior is similar to the exact solutions of the plane-wave collision. 
}
\end{figure}
%%%%%%%%%%%%%%%%%%%%%%%%%%%%%%%%%%%%%%%%%
% Figure <<<<
%%%%%%%%%%%%%%%%%%%%%%%%%%%%%%%%%%%%%%%%%
\end{center}

~

\end{widetext}

%\clearpage

%%%%%%%%%%%%%%%%%%%%%%%%%%%%%%%%%%%%%%%%%
% Figure 8 >>>>
%%%%%%%%%%%%%%%%%%%%%%%%%%%%%%%%%%%%%%%%%
\begin{figure}[ht]
\includegraphics[keepaspectratio=true,width=8cm]{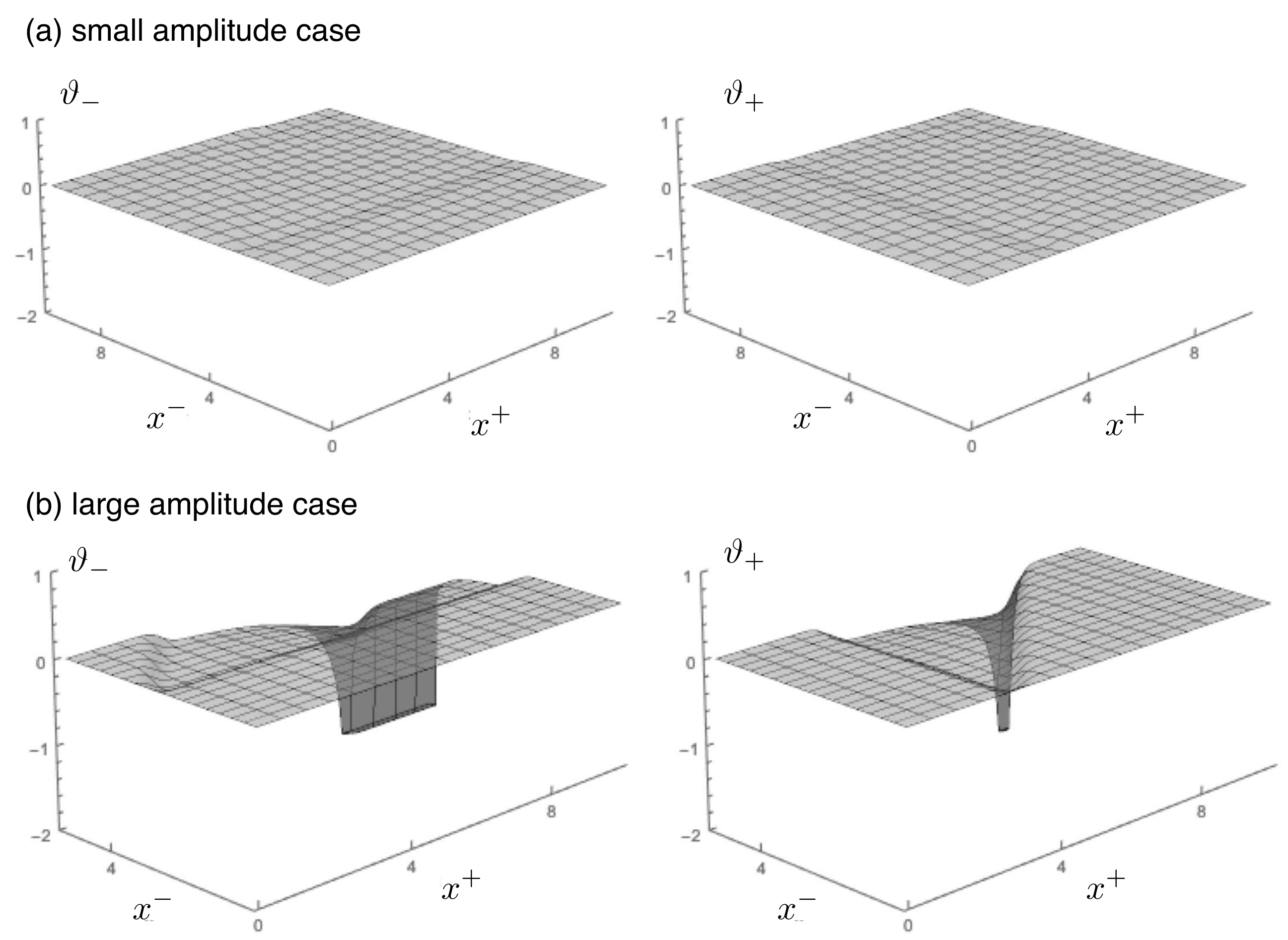}\\
\caption{\label{fig:thetaplusminus}  The expansions $\vartheta_\pm$ for the evolutions shown in Fig.~\ref{fig:SPcollision}. (a) Small-amplitude case ($a=0.2$). (b) Large-amplitude case ($a=0.4$). 
}
\end{figure}
%%%%%%%%%%%%%%%%%%%%%%%%%%%%%%%%%%%%%%%%%
% Figure 8 <<<<
%%%%%%%%%%%%%%%%%%%%%%%%%%%%%%%%%%%%%%%%%

~

~

\subsection{Evolutions in Einstein-GB}
We also evolved the same initial data by the set of evolution equations with nonzero $\alpha_{\rm GB}$. 

\begin{figure}[b]
\includegraphics[keepaspectratio=true,width=8cm]{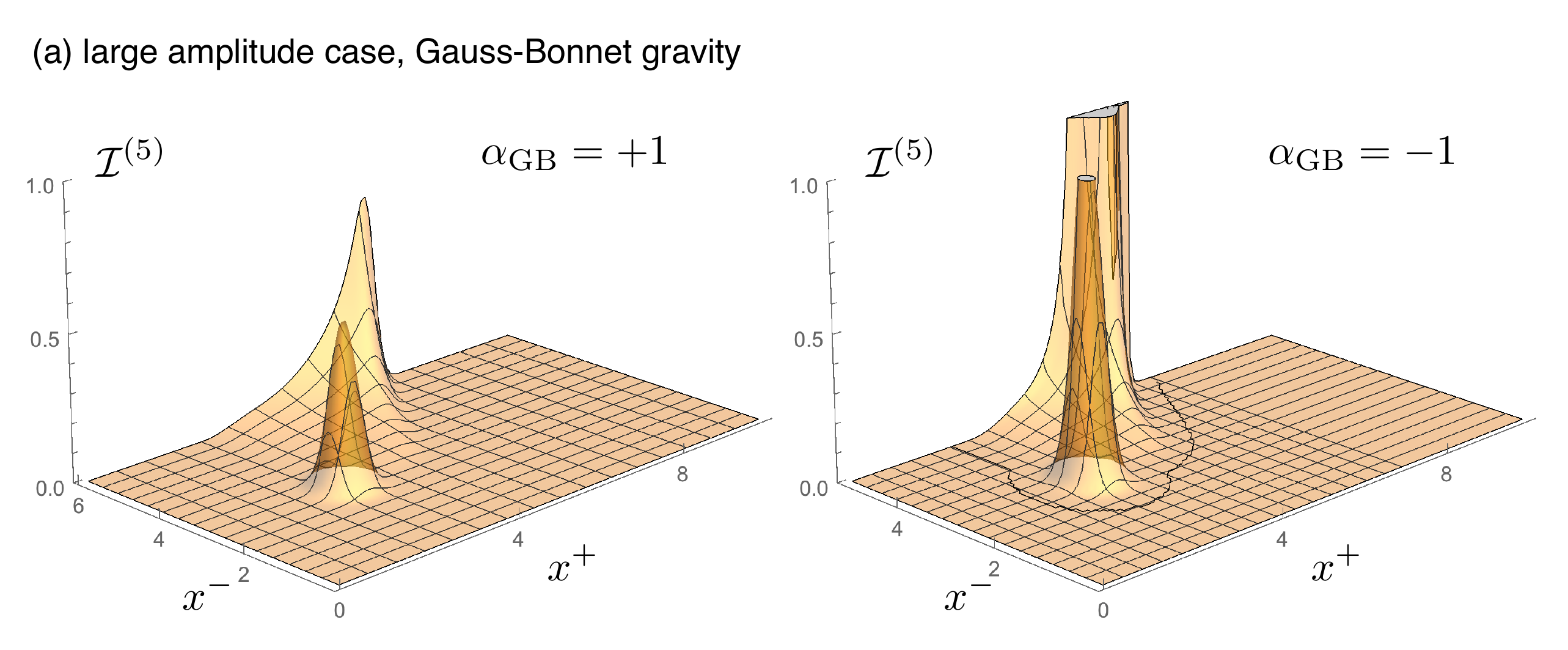}\\
~\vspace{-3cm}\\
\includegraphics[keepaspectratio=true,width=8cm]{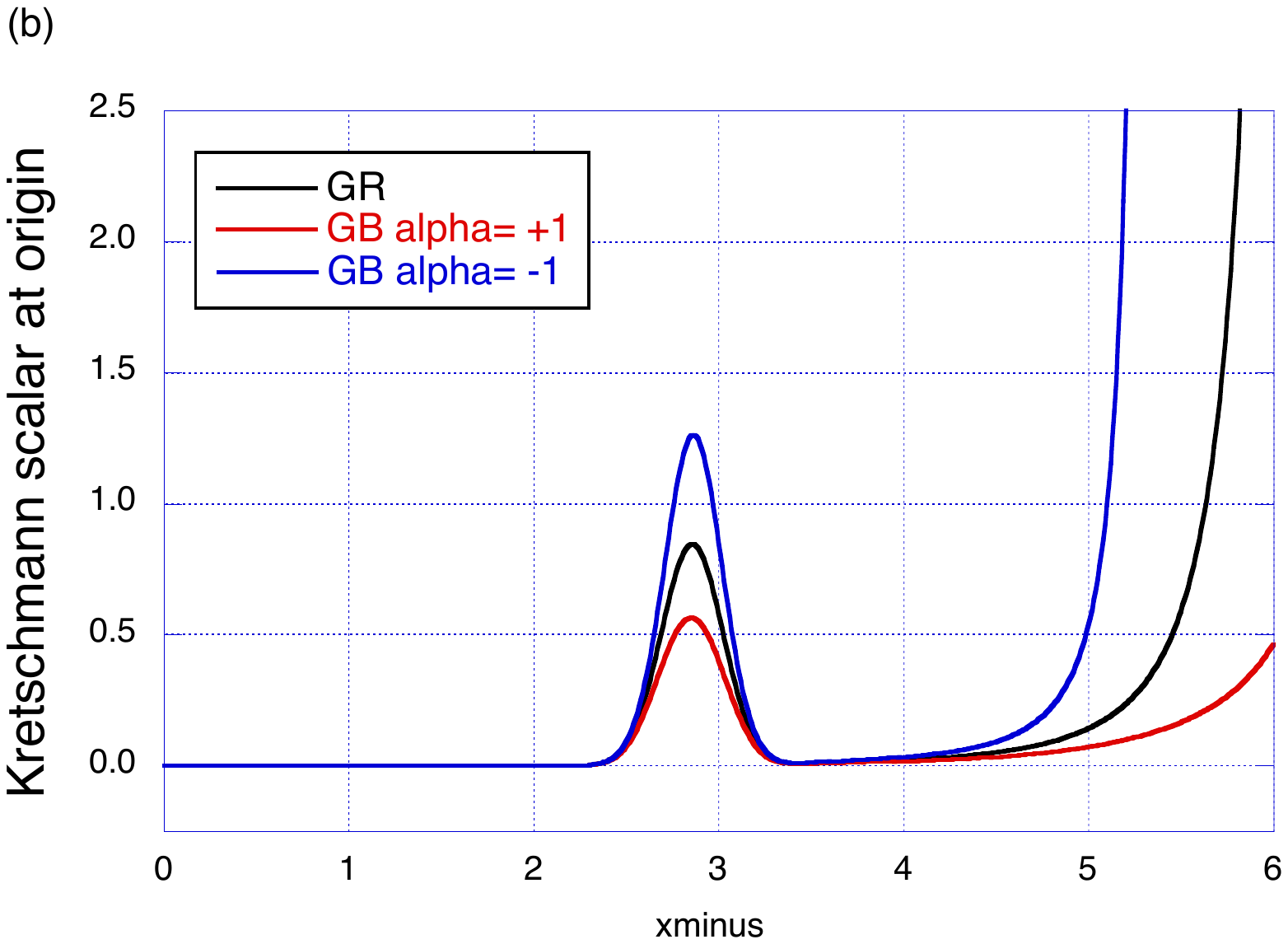}\\
~\vspace{-4cm}
\caption{\label{fig:SPcollisionGB}
(a) Kretschmann scalar, ${\cal I}^{(5)}$, of the evolutions of colliding two scalar pulses in five-dimensional Einstein-GB gravity with $\alpha_{\rm GB}=\pm 1$.  The initial data are the same with the large-amplitude case in Fig.~\ref{fig:SPcollision}(b).  
We see that the local peak of ${\cal I}^{(5)}$ at the collision of two pulses (at $x^+=x^-=2\sqrt{2}$) is smaller (larger) when 
$\alpha_{\rm GB} > 1$ ($\alpha_{\rm GB} < 1$).  (b) Kretschmann scalar, ${\cal I}^{(5)}$, at the origin ($x^+=x^-$), of these evolutions together with one with $\alpha_{\rm GB}=0$ (i.e. GR). 
}
\end{figure}

Figure \ref{fig:SPcollisionGB}(a) displays the Kretschmann scalar, ${\cal I}^{(5)}$, for 
both $\alpha_{\rm GB}=+1$  and $\alpha_{\rm GB}=-1$ cases for the
same initial data with the large-amplitude case ($a=0.4$) in Fig.~\ref{fig:SPcollision}(b). 
We see that the local peak of ${\cal I}^{(5)}$ at the collision of two pulses (at $x^+=x^-=2\sqrt{2}$) is smaller (larger) when 
$\alpha_{\rm GB} > 0$ ($\alpha_{\rm GB} < 0$)  than that in GR. 
This result indicates that introducing the GB terms (in the way of the normal higher-curvature correction; $\alpha_{\rm GB} > 0$) will work for reducing the growth of the local curvature.

%%%%%%%%%%%%%%%%%%%%%%%%%%%%%%%%%%%%%%%%%
% Figure 9 >>>>
%%%%%%%%%%%%%%%%%%%%%%%%%%%%%%%%%%%%%%%%%
%%%%%%%%%%%%%%%%%%%%%%%%%%%%%%%%%%%%%%%%%
% Figure 9 <<<<
%%%%%%%%%%%%%%%%%%%%%%%%%%%%%%%%%%%%%%%%%

%%%%%%%%%%%%%%%%%%%%%%%%%%%%%%%%%%%%%%%%%
% Figure 10 >>>>
%%%%%%%%%%%%%%%%%%%%%%%%%%%%%%%%%%%%%%%%%
\begin{figure}[h]
~\vspace{-3cm}\\
\includegraphics[keepaspectratio=true,width=9cm]{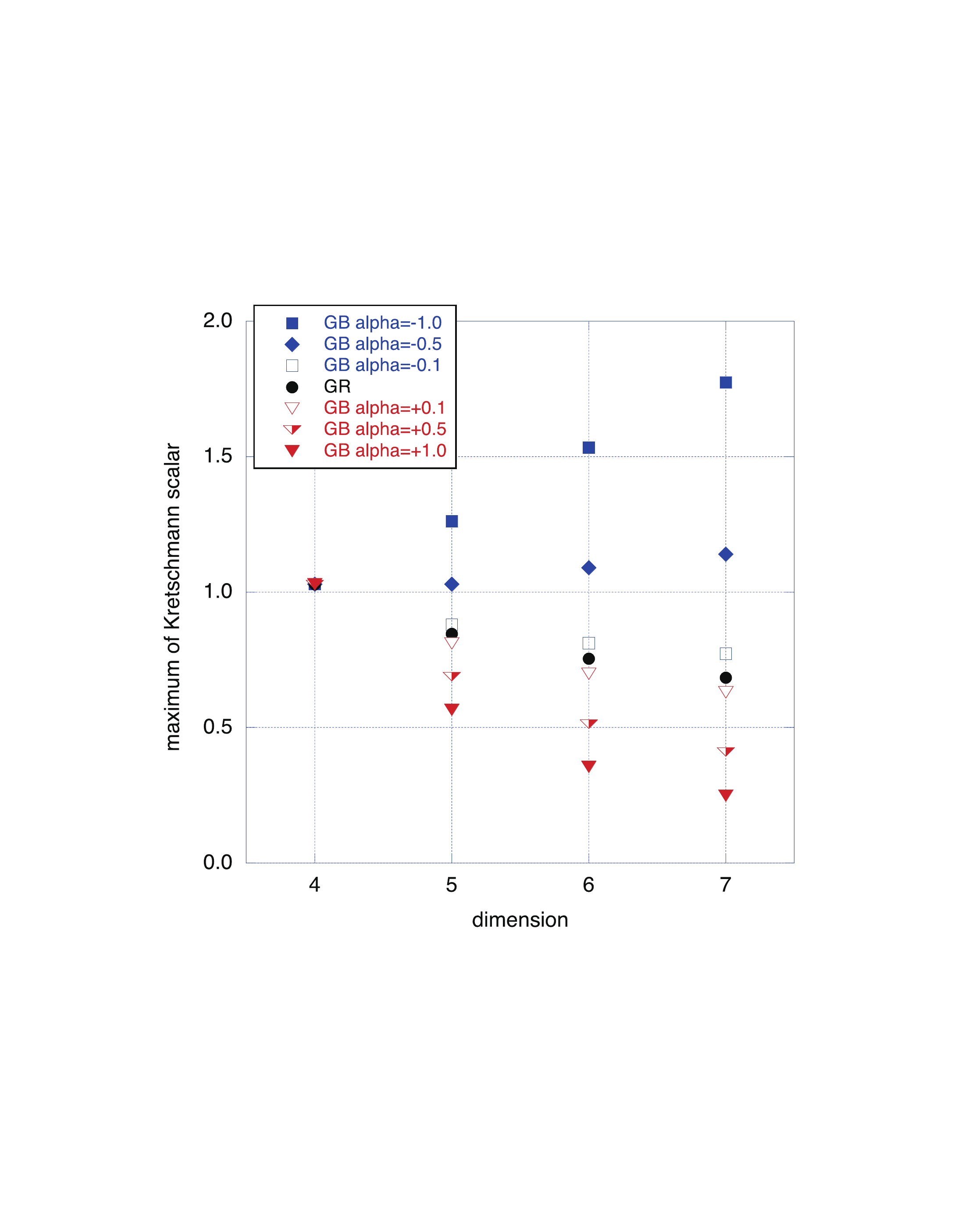}\\
~\vspace{-3cm}
\caption{\label{fig:maxK} The Kretschmann scalar, ${\cal I}^{(n)}$, at the moment of the collision of scalar pulses 
(at $x^+=x^-=2\sqrt{2}$).  We plot for the models with $\alpha_{\rm GB}=0, \pm 0.1, \pm 0.5,$ $ \pm 1.0$ 
  and for the dimensions $n=4,\: 5,\:6$, 7.  
For larger dimensions, the magnitude becomes lower in GR.  We also find that introducing positive $\alpha_{\rm GB}$ (i.e. the normal  higher-curvature correction) reduces its magnitude. 
}
\end{figure}
%%%%%%%%%%%%%%%%%%%%%%%%%%%%%%%%%%%%%%%%%
% Figure 10 <<<<
%%%%%%%%%%%%%%%%%%%%%%%%%%%%%%%%%%%%%%%%%

In Fig.~\ref{fig:SPcollisionGB}(b), we plot the ``evolution" behavior of the Kretschmann scalar, ${\cal I}^{(5)}$, at the origin ($x^+=x^-$) where two pulses collide.  
At later times, we see that the curvature will diverge for all the cases (GR and Einstein-GB) due to the large amplitude of the initial pulses, but these growing behaviors are again ordered by $\alpha_{\rm GB}$.  Supposing that the curvature singularity will be formed at the final phase of this evolution (analogues to the plane-wave collision), then we can say that introducing the GB terms cannot stop the formation of the singularity, but it will shift its appearance later if $\alpha_{\rm GB} > 0$. 

Figure \ref{fig:maxK} shows the magnitude of the Kretschmann scalar, ${\cal I}^{(n)}$, at the moment of the collision of 
scalar pulses (at the first peak of ${\cal I}^{(n)}$). 
  We plot the cases $\alpha_{\rm GB}=0, \pm 0.1, \pm 0.5,$ $ \pm 1.0$ 
  and the dimensions $n=4, 5, 6$, and 7.  We see that for $n=4$, all three cases have the same magnitude, which is consistent with the fact that the GB correction does not appear at $n=4$. For larger dimensions, the magnitude becomes lower.  We also find that introducing positive $\alpha_{\rm GB}$ (i.e. the normal higher-curvature correction) reduces its magnitude.

In summary, the collision of scalar pulses will produce curvature singularity if its initial amplitude is large enough, but its appearance will be delayed in higher dimensions and/or with the GB terms with $\alpha_{\rm GB}>0$.

%######################################################################%
%######################################################################%
%    SECTION  5 
%######################################################################%
%######################################################################%
\section{Summary and Discussions}\label{section5}
%======================================%
%======================================%
The Einstein-GB gravity theory is one of the plausible candidates which describes the early Universe, but so far little is known of its nonlinear dynamical behaviors. 
We numerically investigated the dynamics in higher-dimensional space-time with and without the GB terms. 
We prepared a code for solving the full set of evolution equations in the spherically symmetric or planar symmetric space-time using the dual-null formulation, and we showed the dynamical features on two models, the fate of the perturbed wormhole and the collision of scalar pulses. 
%which is reasonable approach to follow the trapping horizons 

For wormhole dynamics, we monitored the throat structure of the static wormhole by injecting a perturbation to it. 
We confirmed the instability of the Ellis-type wormhole in higher dimensions which was predicted from the linear analysis before. 
We also find that the fate of the wormhole (to either  a black hole or expanding throat) is determined by the signature of the total energy in GR which has the same features as those in four-dimensional cases.  In Einstein-GB gravity, however, we observed that the threshold of the energy which makes a wormhole to a black hole is larger for the  GB correction with the normal sign of the coupling constant ($\alpha_{\rm GB}>0$), and also larger for higher-dimensional cases. These facts indicate that adding the GB terms has similar effects to reducing the total energy of the system. 

For scalar pulses' collision, we observed that curvature (Kretschmann scalar) evolves more mildly in the presence of the normal GB terms  ($\alpha_{\rm GB}>0$) and in higher-dimensional space-time. 
The appearance of the singularity is inevitable in our model, but the basic feature is matched with the expected effect of the cosmologists; i.e., the avoidance (or lower possibility) of the appearance of the singularity. 

Both models suggest the consistent features: the chances of the appearance of a singularity or black hole will be reduced in higher-dimensional space-time and/or in the presence of the GB terms.
As is shown in other models (e.g., Refs. \cite{YamadaShinkai2010CQG,YamadaShinkai2011PRD}), in higher-dimensional GR, the chance of appearances of naked singularities is suppressed compared to the four-dimensional GR cases. This is suggested by the existence of many freedoms in gravity which suppresses the growth of curvature and makes the formation of horizons less eccentric. The introduction of the GB terms seems to work for this direction.   

We hope that these results will be used as a guiding principle for understanding the fundamental dynamical features of the Einstein-GB gravity. 

\section*{Acknowledgments}
This work was supported in part by the Grant-in-Aid for the 
Scientific Research Fund of the JSPS (C) No. 25400277, and also by  
MEXT KAKENHI Grants No. 17H06357 and No. 17H06358.
%Numerical computations were carried out on SR16000 at YITP in Kyoto University, 
%and on the RIKEN Integrated Cluster of Clusters (RICC).

%######################################################################%
%######################################################################%
%    REFERENCES 
%######################################################################%
%######################################################################%

\end{document}